\documentclass[12pt]{article}
\usepackage[round,sort,authoryear]{natbib}

\usepackage{amsmath,amssymb,amsthm,amsfonts, fullpage, setspace}

\usepackage{eucal}
\usepackage{subeqnarray,bm}
 \usepackage[utf8]{inputenc}
\usepackage{graphicx}[final]
\usepackage{psfrag} 
\usepackage{color,float}
\usepackage{pdfpages}

\usepackage{epstopdf}

\usepackage[colorlinks]{hyperref}

\usepackage{hyperref}
\hypersetup{colorlinks=red,citecolor=blue,pdfstartview=FitH, pdfpagemode=None}
\usepackage{multirow}
\usepackage[bottom]{footmisc}
\usepackage{ifthen} 
\usepackage{sectsty}
\usepackage{subfig}
\usepackage{float}
\usepackage[english]{babel}
\usepackage[nottoc]{tocbibind}
\usepackage{calrsfs}
\usepackage{soul}
 \DeclareGraphicsExtensions{.eps,.png,.pdf}
\usepackage{mwe}
\makeatletter
\def\maxwidth{ %
  \ifdim\Gin@nat@width>\linewidth
    \linewidth
  \else
    \Gin@nat@width
  \fi
}
\makeatother

\definecolor{fgcolor}{rgb}{0.345, 0.345, 0.345}

\allsectionsfont{\normalsize\bfseries}

\usepackage{framed}
\makeatletter
 {\par\unskip\endMakeFramed%
 \at@end@of@kframe}
\makeatother

\definecolor{shadecolor}{rgb}{.97, .97, .97}
\definecolor{messagecolor}{rgb}{0, 0, 0}
\definecolor{warningcolor}{rgb}{1, 0, 1}
\definecolor{errorcolor}{rgb}{1, 0, 0}

\newtheorem{thm}{Theorem}[section]

\newtheorem{defn}[thm]{Definition}

\newtheorem{proposition}[thm]{Proposition}

\newtheorem{corollary}[thm]{Corollary}

\newtheorem{remark}[thm]{Remark}

\newcommand{\ds}{\displaystyle}

\newcommand{\set}[1]{\left\{#1\right\}}

\newcommand{\R}{\mathbb{R}}

\newcommand{\p}{\partial}

\newcommand{\E}{\mathbb{E}}

\numberwithin{equation}{section}

\usepackage{calrsfs}

\allsectionsfont{\normalsize\bfseries}

\numberwithin{equation}{section}
\allowdisplaybreaks

\usepackage[table, dvipsnames]{xcolor}
\DeclareMathAlphabet{\pazocal}{OMS}{zplm}{m}{n}
    \newcounter{example}[section]

\begin{document}
\title{
  Information Theory in a Darwinian Evolution Population Dynamics Model
}
\author{Eddy Kwessi\footnote{Corresponding author: Department of Mathematics, Trinity University, 1 Trinity Place, San Antonio, TX 78212, Email: ekwessi@trinity.edu}}
\date{}
\maketitle
\begin{abstract}
Using information theory, we propose an estimation method  for  traits parameters in a  Darwinian evolution model for species with on trait or multiple traits. We  use the Fisher's information to obtain the errors on the estimation  for one species with one or multiple traits. We perform simulations to illustrate the method.
\end{abstract}
\section{Introduction}

Evolution can be thought of as the dynamic changes in  organisms' traits. These changes are mainly due to the environment they live in and occur through mutations or random genetic changes. Darwinian evolution was pioneered by  \cite{Darwin1859} whereby he stated his theory of natural selection as follows: {\it species that are best adapted to their environment would pass their traits onto their offsprings}. This can be expanded to mean that organisms from the same species have the same traits, they may compete for survival, individuals with survival traits may reproduce successfully and pass on their traits to their offsprings, and overtime, organisms that best handle variations may become a separate species. Therefore, tracking overtime variations in species traits may help better understand the evolution of these species. From a theoretical point of view, information theory can help track such changes. Indeed, 
given an observable  random variable $X$ depending on a parameter (or trait) $\theta$, its Fisher's information $I(\theta)$ represents the amount of information that the random variable contains about the parameter (\cite{Lehman1998}). Since its introduction by Fisher (\cite{Fisher1922}), the Fisher's information  has undergone a considerable amount of studies and has been applied to many areas of scientific research. For instance, it has been used to calculate non-informative Jeffrey's priors in Bayesian Statistics (\cite{Bernado1994}), the formulation of Wald's test (\cite{Ward2015}),  it is  connected to the derivation of the Cram\'er-Rao bound of an estimator \cite{Rao1945, Cramer1946}, it is a common feature in optimal design (\cite{Smith1918}), machine learning,  especially elastic weight consolidation (\cite{Kilrkpatrick2017}), color discrimination, \cite{Fonseca2016}, in computational neuroscience to accurately calculate bounds of neural codes (\cite{Abbott1999}). It was recently used in epidemiology to asses  how  two different data sources affect the estimation of the reproductive number of SARS-Cov-2 (\cite{Parag2022}). More importantly,  the Fisher's information is related to information theory via the notion of relative entropy (or information). Indeed, the Fisher's information is the Hessian matrix with respect to the parameter of the relative entropy  or Kullback-Leibler divergence. In more general terms, information can be lost, stored, or gained. In evolutionary biology, information is essentially found in organisms' genomes. It is therefore expected to evolve with changes in the environment. 
Our interest in evolution population dynamics and Fisher's information stems from the work of \cite{Vincent2011}. In it, they discussed Darwinian dynamics and evolutionary game theory. More precisely, a static game with $n$ players is considered where player $i$ chooses his strategy $\theta_i$ to maximize his payoff
\begin{equation}\label{eqn01}
f_i(\Theta),\quad i\in \set{1,2,\cdots, n}\;,
\end{equation}
where $\Theta=(\theta_1,\theta_2,\cdots, \theta_n)$. A dynamic game is then defined as an ordinary differential equation 
\begin{equation}\label{eqn02}
\dot{x}_i=F_i({\bm x},\Theta), \quad \set{1,2,\cdots, n}\;,
\end{equation}
with ${\bm x}=(x_1,x_2,\cdots, x_n)$, where $F_i({\bm x},\Theta)$ is the instantaneous payoff function of player $i$. The static and dynamic games differ in that in the former, the goal is to  maximize a player's payoff whereas in the latter, the goal is to find strategies that persist overtime. Adapting this concept to  ecology in particular, in the presence of $n$ species, the payoff function $f_i(\Theta)$ is referred to as the instantaneous per capita growth rate of a species with density $x_i$ and common  strategy or trait $\theta_i$. To simplify the fitness of many species,  \cite{Vincent2005} introduced the notion of $G$-function as 
\begin{equation}\label{eqn03}
G({\bm x},\theta,\Theta)\big|_{\theta=\theta_i}=f_i({\bm x},\Theta), \quad i \in \set{1,2\cdots, n}\;,
\end{equation}
where $\theta$ is called a ``virtual variable". This leads to an evolutionary equation of strategies given as 
\begin{equation}\label{eqn04}
\dot{\theta}_i=\sigma^2 g({\bm x},\theta, \Theta)\big|_{\theta=\theta_i}\;,\quad \mbox{where $g({\bm x},\theta, \Theta)=\frac{\partial \ln(G({\bm x},\theta, \Theta))}{\partial \theta}$}
\end{equation}
Now consider the evolutionary population dynamical system given as 
\begin{equation}
\begin{aligned}
\dot{x}_i&=x_iG({\bm x},\theta,\Theta)\big|_{\theta=\theta_i}\\
\dot{\theta}_i&=\sigma^2g({\bm x},\theta,\Theta)\big|_{\theta=\theta_i}\;,
\end{aligned}
\end{equation}
where  $\sigma^2$ is the variance of the distribution of strategies (or traits)-values among phenotypes of single species, see \cite{Vincent2011}.
This system has a solution called evolutionary stable strategies (ESS) introduced by \cite{Smith1973} and \cite{Smith1982}. In particular, a necessary condition for the existence of an ESS is that the $G$-function takes on a maximum at zero with respect to $\theta$,  see \cite{Vincent2005}. Since then, there have been many studies of this model, especially its  the discrete version.  In particular, \cite{Ackleh2015} discussed competitive evolutionary dynamics, \cite{Cushing2019} proposed  difference equation schemes for evolutionary population dynamics, which was followed by a new approach in  \cite{Mokni2020}. \cite{Cushing2023} introduced  a Susceptible-Infected (SI) model for a Darwinian model with evolutionary resistance. \cite{Elaydi2022} discussed the effects of evolution on the stability of competing species.  In this paper, among many other things, we would like to address two questions that were raised at the end of the paper by \cite{Vincent2011}, especially in the discrete case, that is:
\begin{enumerate}
\item[(i)] Are there interpretable relations between the maximum of $G$,  $\ds g$,  and $\ds \frac{\partial g}{\partial \theta}<0$?
\item[(ii)] Similarly for $\ds g=0$ and $\ds \frac{\partial g}{\partial \theta}>0$?
\end{enumerate}
Our literature review did not yield meaningful responses to these questions. One way to understand the importance of $G$-functions is in the context of random variables. Indeed, in the one-species case, if $G=G(x,\theta)$ is the density function of a random variable $X$ depending on an unknown trait parameter $\theta$, the amount of information that a random sample from $X$ contains about $\theta$ can be calculated and may lead to the estimation of the unknown  parameter $\theta$. This amount of information or  the Fisher's information (\cite{Lehman1998}) can be calculated theoretically when $G$ is known and may be estimated given an observed random sample. Since most known probability distributions in statistics are from an exponential family, the logarithm of $G$  rather than $G$-itself, its first derivative $\ds g$,  and in some cases its second derivatives $ \ds \frac{\partial g}{\partial \theta }$ can be used to calculate the Fisher's information. While many mathematical questions have been answered and  are being answered in the literature on Darwinian dynamics, computational aspects stil lag behind. Here, we aim to remedy that by providing a statistical framework for  estimating  a species traits in a Darwinian model. More specifically, we will show that minimizing or maximizing  the information (Fisher's information or relative) in a properly defined context may  not only provide new insights  onto the questions above,  but would also enable the experimenter to properly estimate unknown evolution traits parameters, given the data. The rest of the paper is organized as follows: In Section \ref{sec1}, we make a brief overview of Fisher's information theory as it pertains to mathematical statistics. In Section \ref{sec2}, we discuss Fisher's information in conjunction with discrete evolutionary population dynamics.  Finally in Section \ref{sect5}, we make some concluding remarks.
\section{Review of Fisher's Information Theory}\label{sec1}

Let $G(x,\Theta)$ be the density of a random variable $X$, continuous or discrete on an open set $\mathcal{X}\times \Omega\subset \R\times\R^n$. Here $\Theta=(\theta_1, \theta_2, \cdots, \theta_n)$ is either a single parameter or a vector of parameters. We know that since $G(x,\Theta)$ is density function, then $\int_{\mathcal{X}}G(x,\Theta)dx=1$. We also know that  a given  nonnegative integrable function $G_0(x,\Theta)$ defined on $\Omega$ can be made the density of random variable $X$ by considering $G(x,\Theta)=cf_0(x, \Theta)$, where $c^{-1}=\int_{\mathcal{X}}G_0(x,\Theta)dx$.\\
In the sequel, we will always make the following assumptions on the function $G(x,\Theta)$:
\begin{enumerate}
\item[$A_1:$] The support $\set{x\in \mathcal{X}: G(x,\Theta)\neq 0}$ of $G$ is independent of $\Theta$.
\item[$A_2:$] $G(\cdot, \Theta)$ is nonnegative for all $\Theta\in \Omega$ and $G\in L^1(\mathcal{X})$.
\item[$A_3:$] $G(x,\cdot)\in C^2(\Omega)$, the set of continuously and twice differentiable  functions of $\Theta$, for all $x\in \R$.
\end{enumerate}
The first assumption puts out of consideration any  uniform distribution $G(x,\theta)=\frac{1}{\theta}$,  whose support is the interval $(0,\theta)$. The second assumption allows the well-definiteness of $\lambda(x,\theta)=\ln(G(x,\theta))$, its first derivative or score function $g(x,\Theta)=\nabla_{\theta} \lambda(x,\Theta)$ and its second derivative $h(x,\Theta)=\nabla_{\Theta}g(x,\Theta)=\nabla_{\Theta}^2\lambda(x,\Theta)$. We will denote the expected value of a random variable $X$ as $\E[X]$.
\begin{defn}\label{def1}
Given a random variable $X$ with density function $G(x,\Theta)$ satisfying $A_1$ and $A_2$, the Fisher's information of $X$ is defined as 
\begin{equation}\label{eqn0}
I(\Theta)=\E_{X}[(g(X,\Theta))^2]=-\E_{X}[h(X,\Theta)]\;.
\end{equation}
\end{defn}
\noindent When $\Theta$ is a vector of more than one coordinates, the Fisher's information is a symmetric positive definite (thus invertible) matrix $I(\Theta)=(I_{kl}(\Theta))_{1\leq k,l\leq n}$, where 
\begin{equation}
I_{kl}=\E_X\left[\frac{\p^2\lambda(X,\Theta)}{\p \theta_k \p \theta_l}\right], \quad \mbox{for $1\leq k,l\leq n$}\;.
\end{equation}
 The Fisher's information $I(\Theta)$ represents the amount of information contained in an estimator of $\Theta$, given data $X$. In that regard, it is also known as the {\bf observed information} about $\Theta$ contained in an estimator of $\Theta$.  In fact, if $X_1, X_2, \cdots, X_n$ is  a random sample from the distribution $G(x, \Theta)$, the Fisher's information contained in an estimator of $\Theta$, given the  data $X_1, X_2, \cdots,X_n$ is 
\[I_n(\Theta)=nI(\Theta)\;.\]
We recall the following basic definitions:
\begin{defn} Let $X$ be a random variable depending on a vector of parameter $\Theta$ and let $X_1, X_2,\cdots, X_n$ be a random sample generated from $X$. Then 
\begin{itemize}
\item $T=T(\Theta)$ is called an estimator of $\Theta$ based on the random sample $X_1, X_2,\cdots, X_n$ if $T$ is a function of $X_1, X_2,\cdots, X_n$, that is, $T=T(X_1, X_2,\cdots, X_n)$.
\item If $T$ is an estimator of $\Theta$, $T$ is called unbiased if $\mathbb{E}[T]=\Theta$.
\item An unbiased estimator $T$ is called efficient if $\mbox{Var}(T)=\frac{1}{I_n(\Theta)}$.
\end{itemize}
\end{defn}
\noindent In particular, if $T=T(\Theta)$ is an estimator of $\Theta$ based on a sample $X_1, X_2, \cdots, X_n$, the Cram\'{e}r-Rao (see  for instance \cite{Rao1945, Cramer1946}) bound gives an estimate of the best lower bound  for the variance of $T$ as:
\begin{equation}\label{eqn1}
\mbox{Var}(T)\geq (\nabla_{\Theta}\E(T))^TI_n(\Theta)^{-1}(\nabla_{\Theta}\E(T))\;.
\end{equation}
Equality is obtained in \eqref{eqn1} if $T$ is  efficient. If $T$ is unbiased, then $\E[T]=\Theta$ and consequently, $\nabla_{\Theta}\E[T]={\bf 1}$, where ${\bf 1}$ is a vector of ones $\R^n$. Therefore, \eqref{eqn1} becomes
\begin{equation}\label{eqn2}
\mbox{Var}(T)= I_n(\Theta)^{-1}\;.
\end{equation}
{\bf Example 1:} Suppose $X$ is a random variable with distribution $f(x,\theta)=\theta e^{-\theta x}$, where $\theta>0$. Then we have that $\lambda(x,\theta)=\ln(\theta)-\theta x, g(x,\theta)=\frac{1}{\theta}-x$, and $h(x,\theta)=-\frac{1}{\theta^2}$.  It follows that $I(\theta)=-\E_X[h(x,\theta)]=\frac{1}{\theta^2}$ and $I_n(\theta)=\frac{n}{\theta^2}$. If we now consider a random sample $X_1, X_2, \cdots, X_n$ from $X$, and $\ds T=T(\theta)=\frac{1}{n}\sum_{i=1}^n X_i$, we note that $\E[X_i]=\theta$, so that $\E[T]=\theta$. Thus, $T$ is an unbiased estimator of $\theta$ and $\frac{d}{d\theta}T=1$. Moreover, $\mbox{Var}(T)=\frac{\theta^2}{n}$. We then verify  from \eqref{eqn2} that  indeed we have $\mbox{Var}(T)=\frac{\theta^2}{n}=\frac{1}{I_n(\theta)}$\;. Consequently, accurate estimates of $\Theta$ have large Fisher's information (matrix) components  whereas inaccurate ones have small Fisher's information components. 

\section{Evolution population dynamics and  information theory}\label{sec2}

\subsection{Single population model with one trait}
Consider the following discrete  evolutionary dynamical  model 

\begin{equation} \label{eqn3}
\begin{cases}
x_{t+1}& \vspace{0.1cm}=x_tG(x_t, \theta_t, u_t)\\
\theta_{t+1}&=\theta_t+\sigma^2 g(x_t, \theta_t,u_t)
\end{cases}\;,
\end{equation}
where $G(x,\theta, u)=b(\theta)e^{-c_u(\theta)x}$  where $b(\theta)=b_0e^{-\frac{\theta^2}{2w^2}}$ and $c_u(\theta)=c_0e^{-\kappa(\theta-u)}$, for a constant $u$ and for some positive constants $\sigma$ (speed of evolution), $b_0$ (initial birth rate), $c_0$ (competition  constant), $\kappa$, and $w$ (standard deviation of the distribution of birth rates), and  for a differentiable function $c_u(\theta)$ of $\theta$ and positive and continuous function $b(\theta)$. This system  has nontrivial fixed points $(x^*, \theta^*)$ if they satisfy  the equations 
\begin{equation}\label{eqn:OneTrait}
\begin{cases}1&=b(\theta)e^{-c_u(\theta)x}\\
 \frac{1}{b(\theta)}&=c_u'(\theta)x.
 \end{cases} 
 \end{equation}
  This can further be reduced to the condition on $b(\theta)$ and $c_u(\theta)$ given by \begin{equation}\label{Cond}
\frac{\ln(b(\theta))}{c_u(\theta)}=\frac{1}{b(\theta)c_u'(\theta)}\;. \end{equation} 
The theorem below shows how to obtain the Fisher's information of the above system as a function of the system's parameters. 
\begin{thm} \label{thm:FishOneSpecOneTrait}
Let $\ds \Gamma_u(\theta)=\frac{b(\theta)}{c_u(\theta)}$. Then the Fisher's information of this system is constant and  given by 
\begin{equation}\label{eqn:FishInfo}
I(\theta)=\frac{1}{\omega^2}+\kappa^2\Gamma_u(\theta)\;.
\end{equation}
\end{thm}
\noindent The proof can be found in Appendix $A_1$.\\
The Corollary below shows that the Fisher's information has a maximum value. 
\begin{corollary} \label{cor:FishOneSpecOneTrait}
The Fisher's information $I(\theta)$ attains its maximum value $I_{\mbox{max}}$ for $\theta_{\mbox{max}}=w^2\kappa$ and the maximum value is  
\begin{equation}\label{eqn:FishInfoMax}
I_{\mbox{max}}= \frac{1}{w^2}+\kappa^2 \frac{b_0}{c_0}e^{\frac{1}{2}(w\kappa)^2-\kappa u}\;.
\end{equation}
\end{corollary}
\noindent The proof can be found in Appendix $A_2$.\\
\begin{figure}[H] 
   \centering
   \includegraphics[scale=0.8]{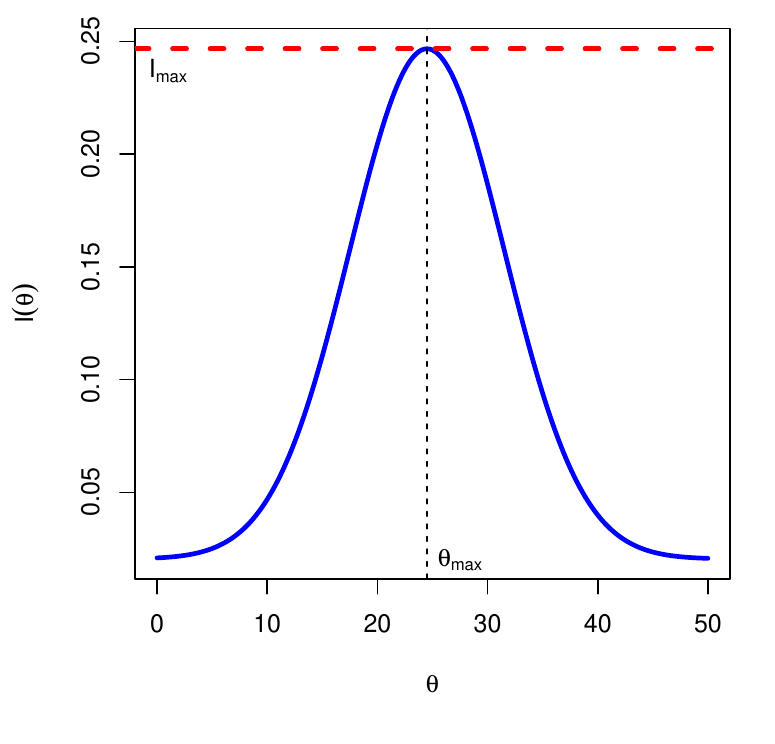} 
   \caption{The blue curve represents the Fisher's Information in equation \eqref{eqn:FishInfo} with its maximum value represented by the red dashed line, for $w=7; \kappa_1=0.5; u=0.02; \frac{b_0}{c_0}=10^{-4}$.}
   \label{fig:FishMax}
\end{figure}
\noindent In the proposition below, we give precise conditions for the existence of nontrivial fixed points of the system above. 
\begin{proposition}\label{prop:OnTraitfixedPoints}
Let \[\xi_1=\frac{1}{\kappa^2}+2w^2\ln(b_0)\;.\] 
If $\xi_1<0$, then the Darwinian system \eqref{eqn3} does not have a nontrivial critical point.\\
\noindent If $\xi_1=0$,  then the Darwinian system has a unique nontrivial fixed point $(x_*,\theta_*)$ given as 
\[x_*=\frac{\ln(b_0)-\frac{\theta_*}{2w^2}}{c_0e^{-\kappa(\theta_*-u)}},\quad \theta_*=-\frac{1}{\kappa}\;.\]
If $\xi_1>0$, then the Darwinian system has two non trivial fixed points $(x_{*+},\theta_{*+})$ and $(x_{*-},\theta_{*-})$ given by 
\[ x_{*\pm}=\frac{\ln(b_0)-\frac{\theta_{*\pm}}{2w^2}}{c_0e^{-\kappa(\theta_{*\pm}-u)}}, \quad \theta_{*\pm}=-\frac{1}{\kappa}\pm \sqrt{\xi_0}\;.\]
\end{proposition}

\noindent Proposition \ref{prop:OnTraitfixedPoints} and Theorem \ref{thm:FishOneSpecOneTrait} are important in that of when $\theta_t\to \theta_{*}$ as $t \to \infty$, then by continuity of the function $I(\theta)$ with respect to $\theta$,  we will have $I(\theta_t)\to I(\theta_*)$ as $t \to \infty$. This means that the Fisher's information, overtime,  will be maximized at the critical point $(x_*,\theta_*)$ of the dynamical system. Therefore, for estimation purposes, the reciprocal of  $I(\theta_*)$  will be the smallest variance for any unbiased estimator of the  trait $\theta$. In Figure \ref{fig:FixedPoints1DFishInfo} below, we use the following parameters: $w=3; \kappa=3; b_0=10; c_0=0.5;  u=1, \sigma=5, n=750, x_0=1, \theta_0=10$.

\begin{figure}[H] 
   \centering
   \centering
   \begin{tabular}{cc}
   {\bf (a)} & {\bf (b)}\\
      \includegraphics[scale=0.6]{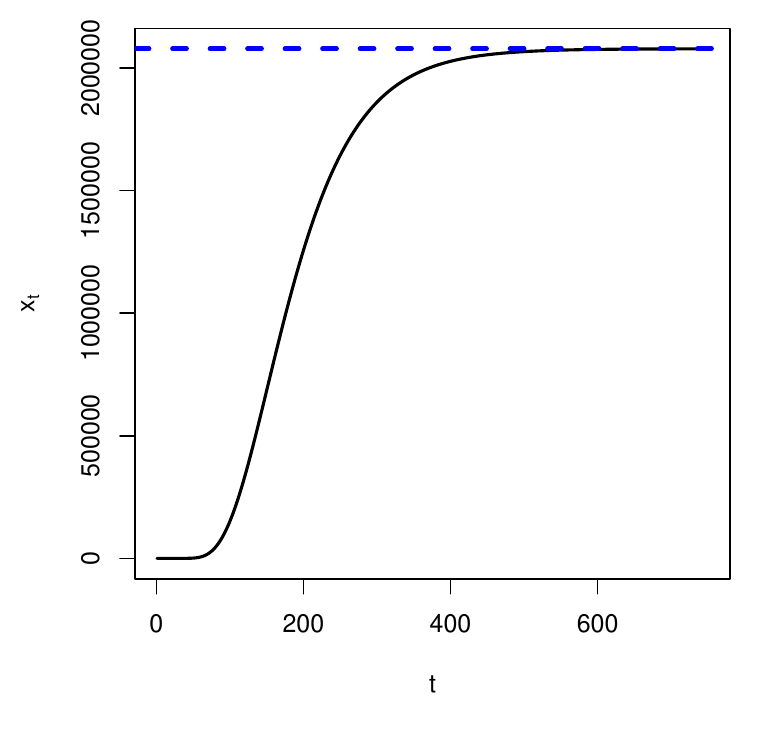} & \includegraphics[scale=0.6]{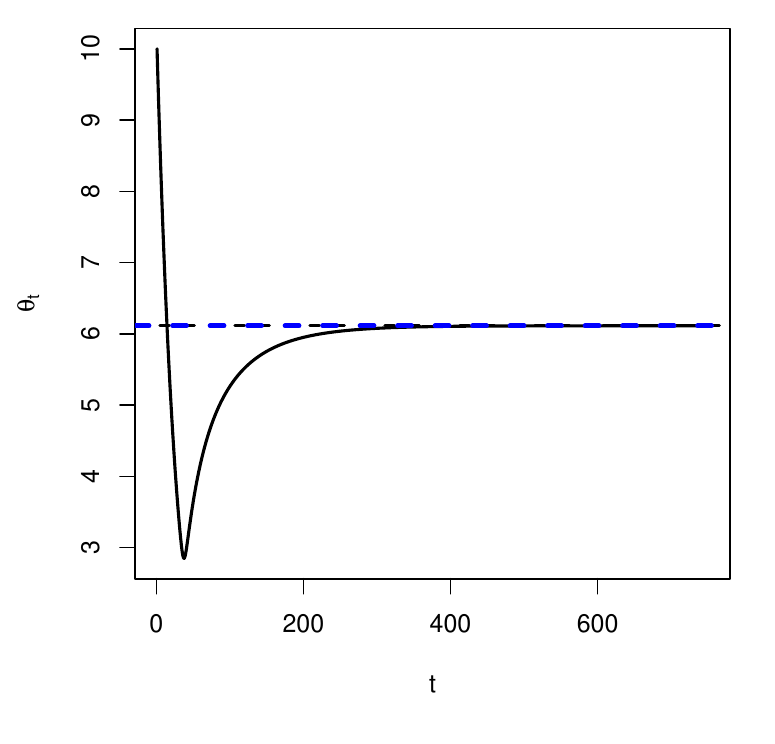}\\
      {\bf (c)} & {\bf (d)}\\
        \includegraphics[scale=0.6]{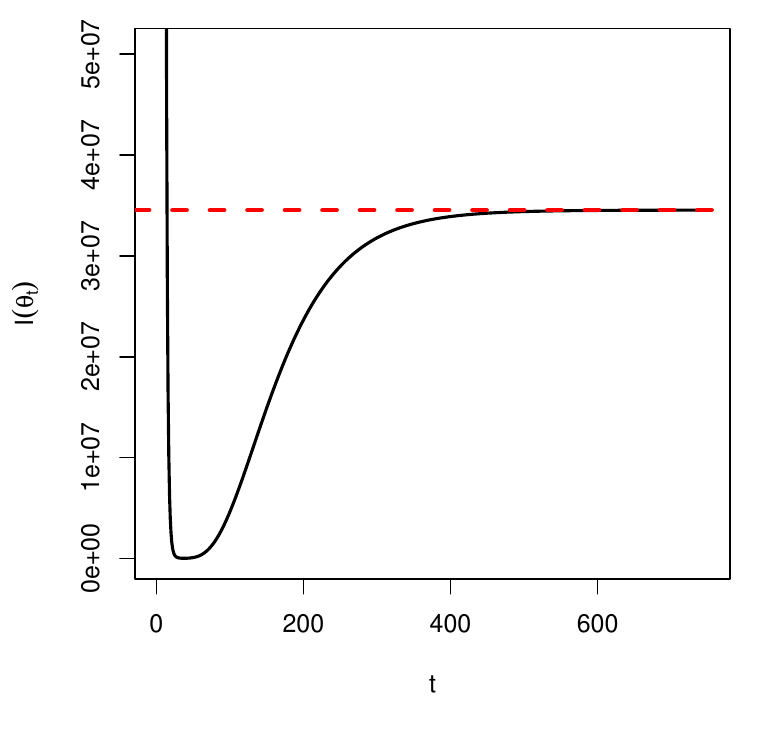} & \includegraphics[scale=0.6]{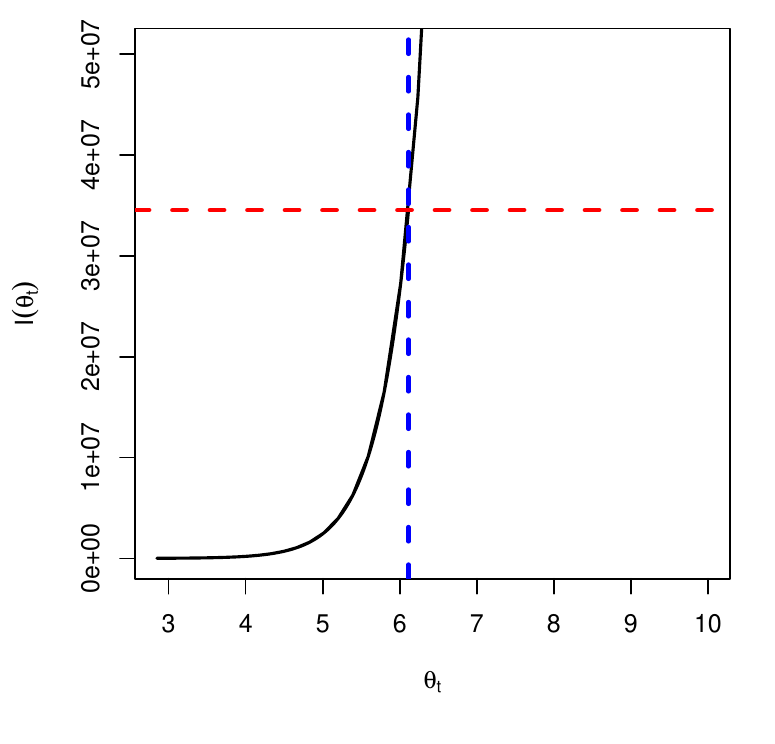}
   \end{tabular}

   \caption{In {\bf (a)}, represented is the time series of of $x_t$. It shows a convergence to $x^*\approx 20.789\times 10^5$ (blue dashed line). In {\bf (b)}, represented is the time series of $\theta_t$, showing a convergence to $\theta_{*+}\approx 6.113$ (blue dashed line). Figure {\bf (c)} represents the time series of the Fisher's information $I(\theta_t)$, showing a convergence to $I(\theta_{*+})\approx 345.43\times 10^5$ (red dashed line). Figure {\bf (d)} is the plot of $I(\theta_t)$ versus $\theta_t$, showing that once the fixed point $\theta_{*+}$ is reached, the Fisher's information is maximized. This is illustrated by the intersection between the blue and red dashed lines. }
   \label{fig:FixedPoints1DFishInfo}
\end{figure}

\noindent  {\bf Special case:} \mbox{}\\

\noindent We will now discuss the particular case of an exponential distribution, that is, $b(\theta)=c_u(\theta)=\theta$. Clearly, the condition  \eqref{Cond} is satisfied with $\theta=\theta_*=:e$ and $x=x_*:=e^{-1}$. Therefore 
\begin{equation} 
\begin{aligned}
G(x,\theta)&=\theta e^{-\theta x}\\
 \lambda(x,\theta)&=\ln(\theta)-\theta x\\
  g(x,\theta)&=\frac{1}{\theta}-x,\\
   h(x,\theta)&=-\frac{1}{\theta^2}\;.
   \end{aligned}
   \end{equation}
    This implies that $\ds I(\theta)=-\E_X[h(X,\theta)]=\frac{1}{\theta^2}$. 

    It follows that there are equilibrium  fixed points: the extinction equilibrium (trivial point) $E_{0}=(0,0)$ and the  interior equilibrium (nontrivial point )$E_{1}=(e^{-1},e)$. In Figure \ref{fig1} below, we represent   functions  $G(x,\theta), F(x,\theta=xG(x,\theta)$, and $\lambda (x,\theta)$ for $\theta=\frac{3}{2}$. This shows that $\lambda(x,\theta)$ is minimized where $G(x,\theta)$ is maximized, providing a clue as to the relation between the maximum of $G$ and the critical points of $g$ and $\frac{\partial g}{\partial \theta}$. Another clue can be found in Figure \ref{figFishInfo1D} below. 
\begin{figure}[H] 
   \centering
   \includegraphics[scale=0.8]{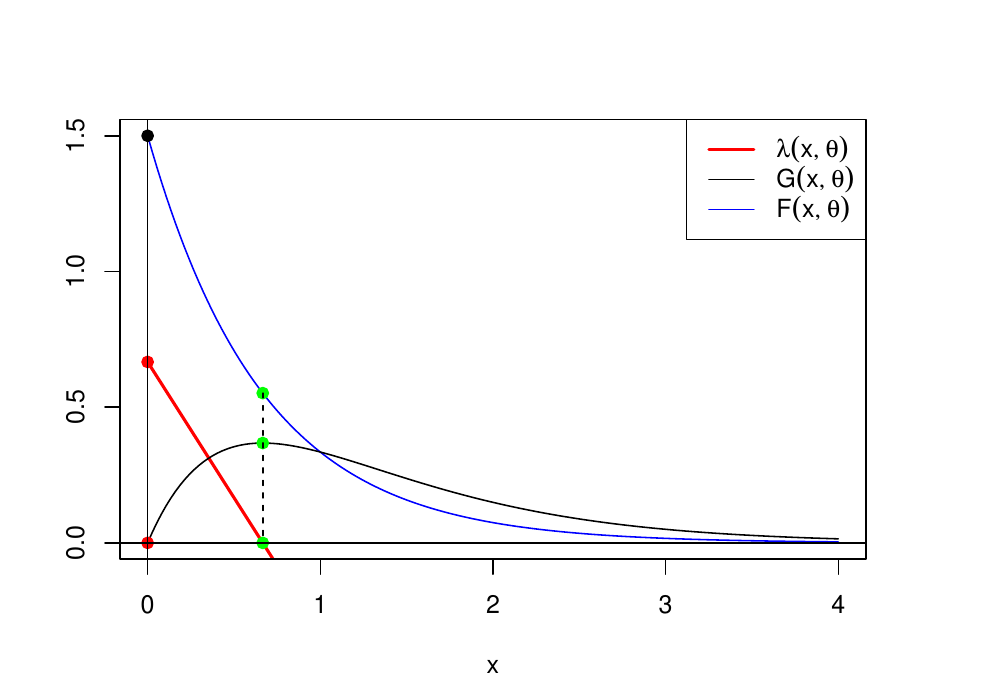} 
   \caption{This figure shows $F(x,\theta):=xG(x,\theta)$ in blue, $G(x,\theta)$ in black and $\lambda(x,\theta)$ in red for $\theta=1.5$. The green dots represent the intersection between the vertical $\ds x=\frac{1}{\theta}=\frac{2}{3}$ and these curves.  We observe that $G(x,\theta)$ is maximized at the same point $x$ where $\lambda(x,\theta)$ is minimized (green dots) and vice versa (red dots).  }
   \label{fig1}
\end{figure}
\begin{figure}[H] 
   \centering
   \centering
   \begin{tabular}{cc}
   {\bf (a)} & {\bf (b)}\\
      \includegraphics[scale=0.6]{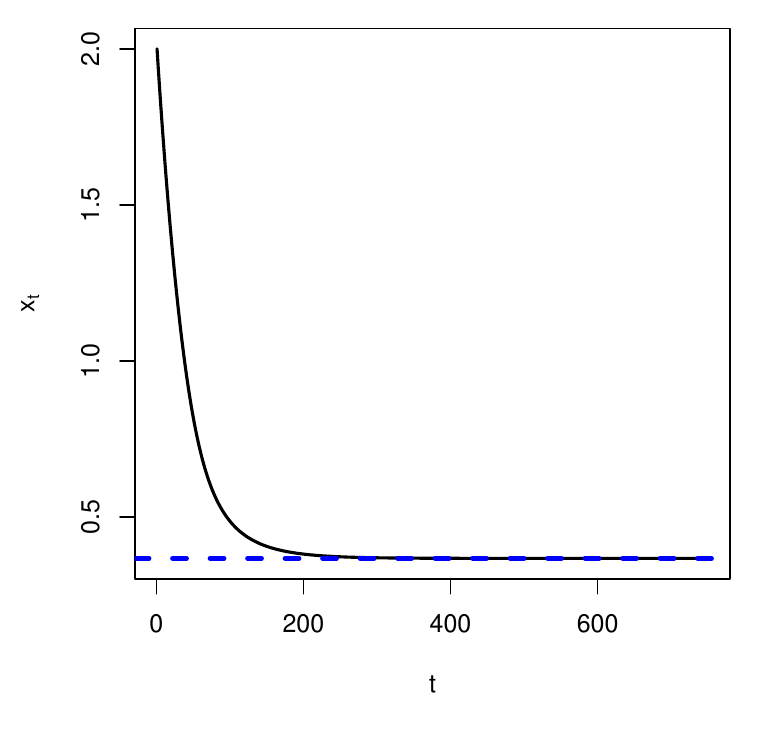} & \includegraphics[scale=0.6]{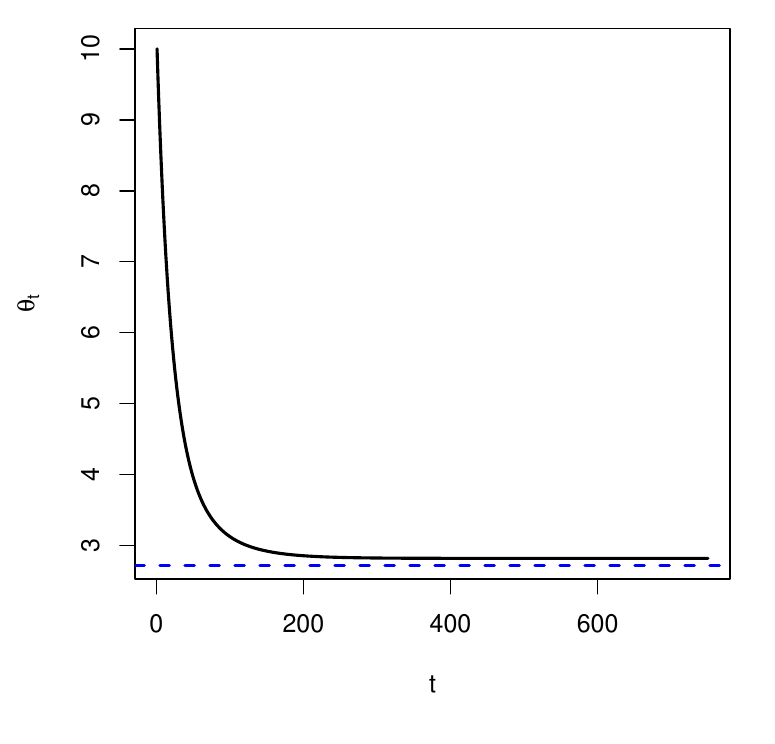}\\
      {\bf (c)} & {\bf (d)}\\
        \includegraphics[scale=0.6]{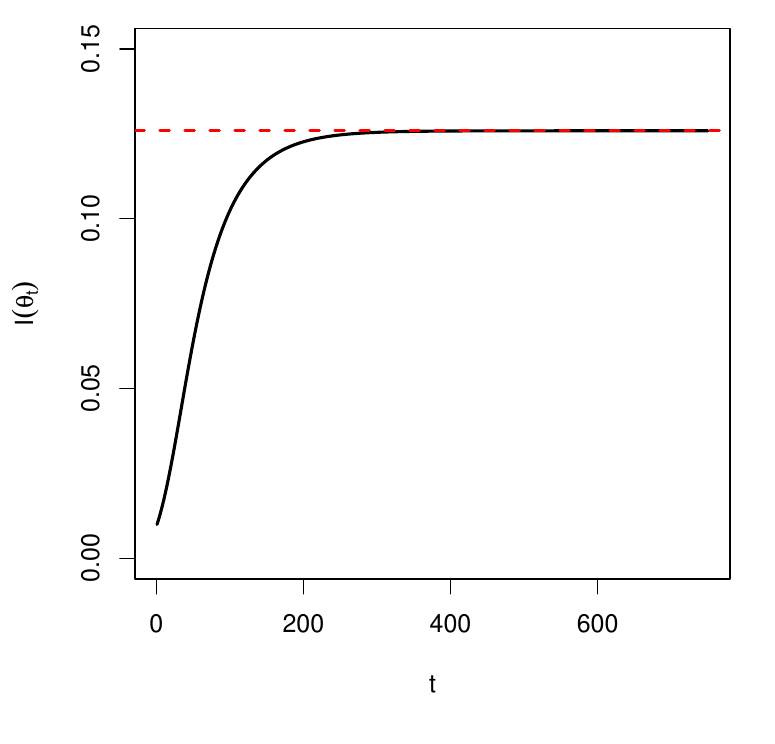} & \includegraphics[scale=0.6]{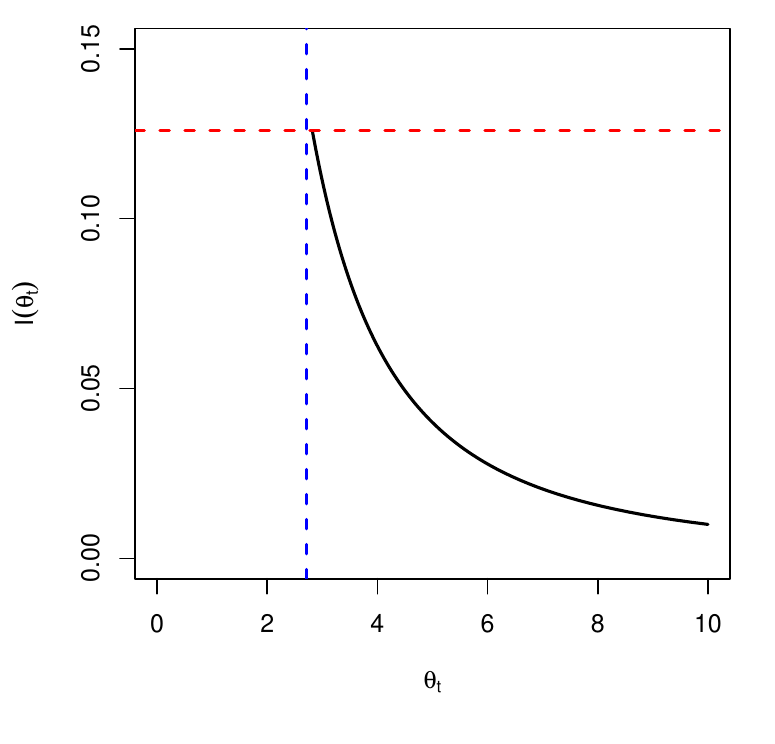}
   \end{tabular}

   \caption{In {\bf (a)}, represented is the time series of of $x_t$ in the special case above. It shows a convergence to $x^*=e^{-1}$ (blue dashed line). In {\bf (b)}, represented is the time series of $\theta_t$, showing a convergence to $\theta_{*+}=e$ (blue dashed line). Figure {\bf (c)} represents the time series of the Fisher's information $I_n(\theta_t)$, showing a convergence to $I(\theta_{*})\approx 0.125$ (red dashed line) as $t\to \infty$. Figure {\bf (d)} is the plot of $I(\theta_t)$ versus $\theta_t$, showing that once the fixed point $\theta_{*+}=e$ is reached, the Fisher's information ix maximized. This is illustrated by the intersection between the blue and red dashed lines. }
   \label{figFishInfo1D}
\end{figure}
\noindent  From a dynamical systems' perspective, it means that $X_t$, the value of $X$ at time $t$,  is generated from  the distribution $G(x,\theta)$ and used to calculate the value of $X_{t+1}$. Therefore, the  role of the first choice of  $\theta$ is to initialize the dynamical system. Once the system is initialized for $X_t$, we can use an information theory approach to provide an estimator $T(\theta)$ of $\theta_t$, the value of $\theta$ at time $t$. That estimator, if efficient,  will have variance $I_n(\theta_t)^{-1}$. We can then use the dynamical system \eqref{eqn3} to estimate $\theta_{t+1}$, and the Fisher's information will provide its variance. We note that $\theta_{t+1}$ will only be an estimate of the true value and therefore will carry an error as $t$ changes. It is therefore expected that at the non trivial critical point (ESS) $(x^*,\theta^*)$ of the dynamical system, the estimator $T(\theta)$ converges to $\theta^*$ and the variance of $T(\theta)$ converges to $I_n(\theta_*)^{-1}$ as $t\to \infty$. \\
\noindent Figure \ref{fig2} below is an illustration of this fact for $G(x,\theta)=\theta e^{-\theta x}$.  
\begin{figure}[H] 
   \centering
   \centering 
   \begin{tabular}{cc}
   {\bf (a)} & {\bf (b)}\\
  \includegraphics[scale=0.58]{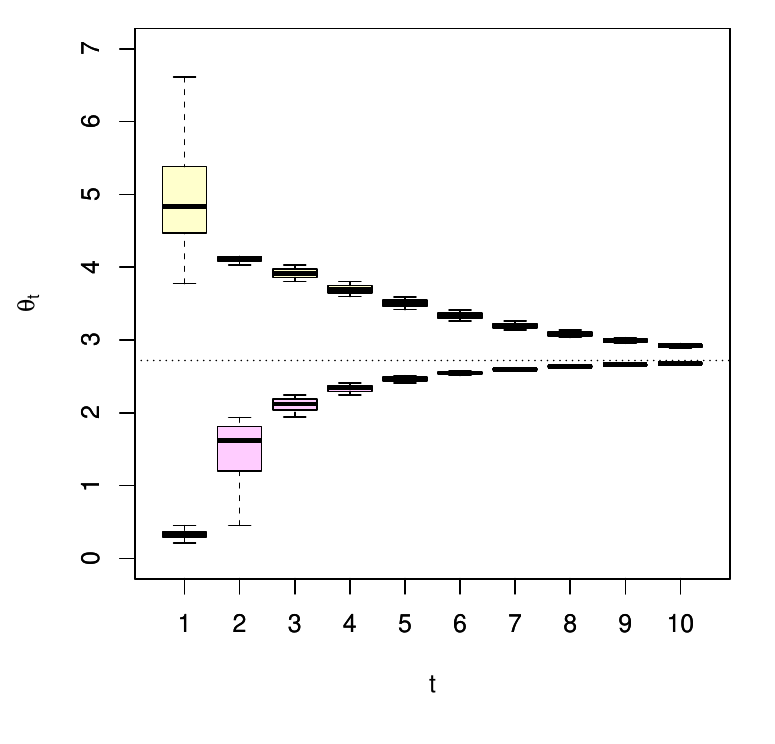} &  \includegraphics[scale=0.58]{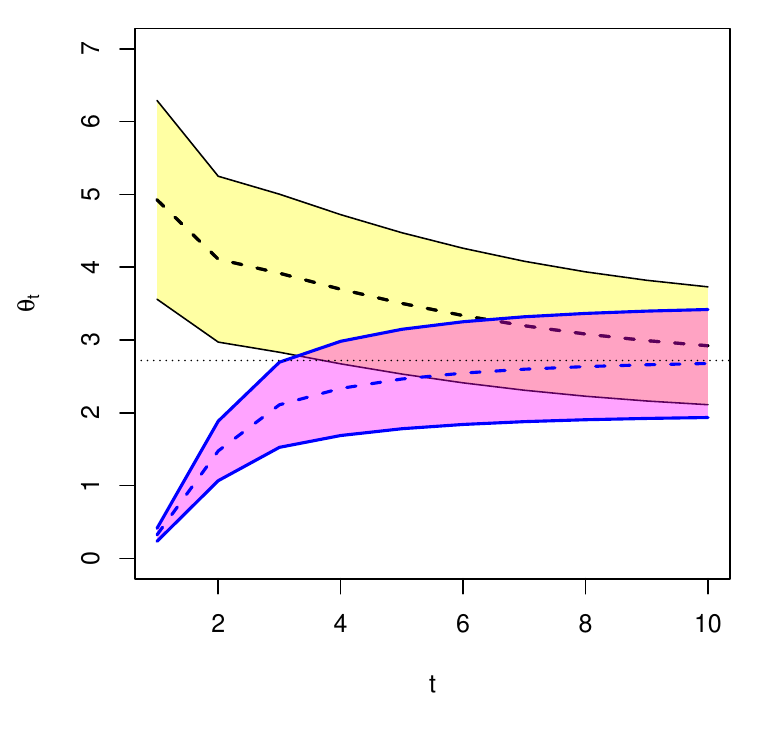} \\
     {\bf (c)} & {\bf (d)}\\ 
     \includegraphics[scale=0.58]{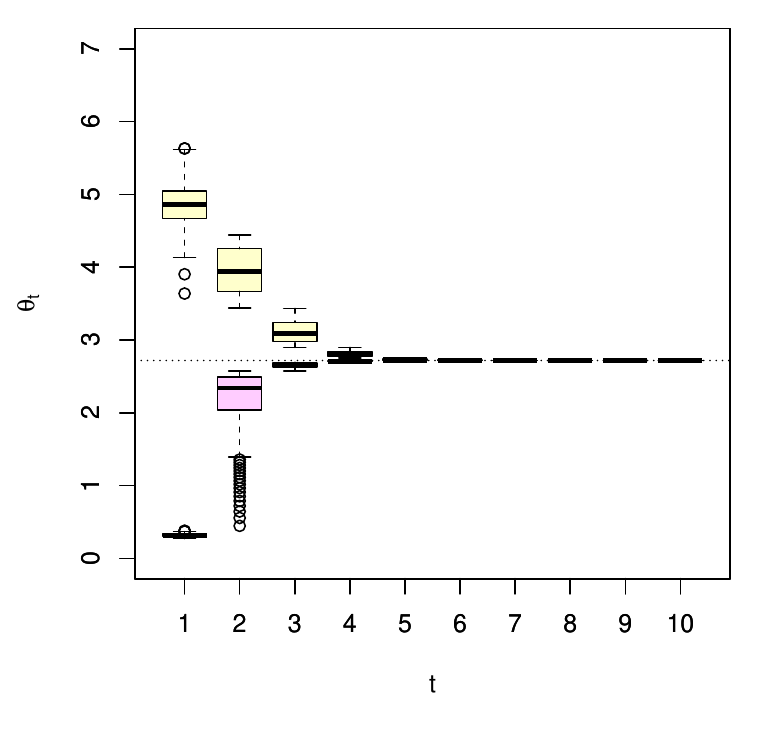} &  \includegraphics[scale=0.58]{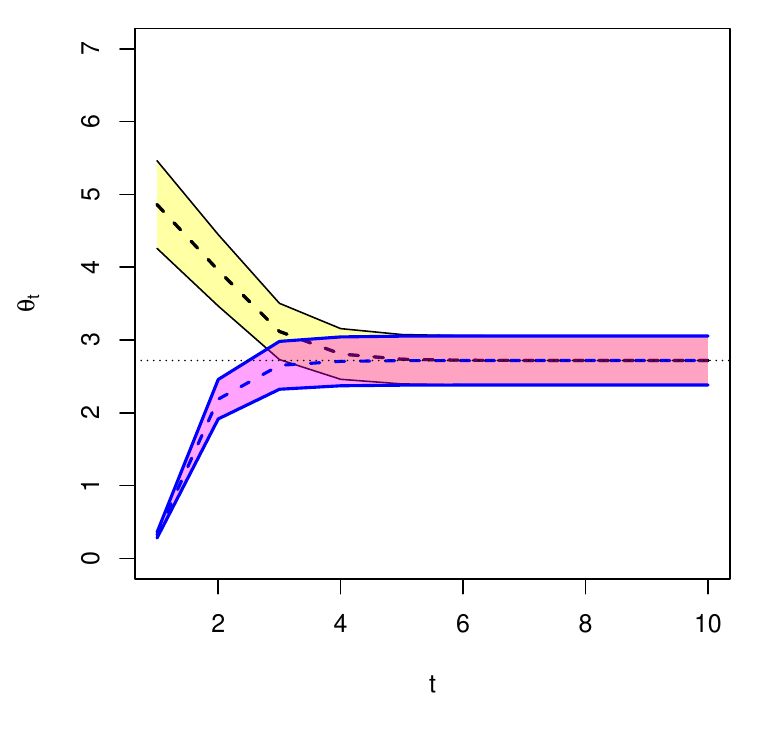}
  \end{tabular}
   \caption{Values of $T_n(\theta_t)$ (dashed line) for two different starting values of $\theta$, each with their 95\% confidence bands (colored shaded areas). In each cases, $T_n(\theta_t)$ converges to $e$ (light dashed line) as $t\to \infty$. When $n$ is large as in {\bf (c)} and {\bf (d)}, $I_n(\theta)^{-1}$ becomes smaller and so is the width of the confidence interval. }
   \label{fig2}
\end{figure}
\noindent Indeed, we  generated two random samples of size $n=50$ (Figure \ref{fig2} {\bf (a)} and {\bf (b)}) and $n=250$ (Figure \ref{fig2} {\bf (c)} and {\bf (d)}) with $t=1,2\cdots, m$ where $m=10$ and  $\sigma=0.04$  from exponential distributions with respective initial parameters $\theta_0=0.2, 2.13$. We choose $\ds T_n (\theta_t)=\frac{1}{n}\sum_{i=1}^nX_i$ which is known to be an efficient estimator of $\theta$. The dashed lines  represent the respective values of $T_n(\theta_t)$, and the black lines represents their 95\% confidence intervals $(T_n (\theta_t)-1.96I_n(\theta_t)^{-1}, T_n (\theta_t)+1.96I_n(\theta_t)^{-1})$. This shows in particular that on average $T_n(\theta_t)$ converges to  $e$ (dashed line), the fixed point of the dynamical system as expected from  above.  
It also shows  that as the Fisher's formation gets larger,  the variance of the estimator gets smaller and thus  the width of confidence interval gets smaller and quickly approaches zero as in  Figure \ref{fig2} {\bf (c)} and {\bf (d)}. 

\begin{remark}
We observe that convergence of $\theta_t$ towards $e$ as predicted depends on choosing appropriate value of $\sigma$. Large values of $\sigma$ will definitely make the system unstable as oscillations will slowly and increasingly occur, see Figure \ref{fig:FixPointsOscill} below. 
\end{remark}
\begin{figure}[H] 
   \centering
   \begin{tabular}{cc}
   \includegraphics[scale=0.6]{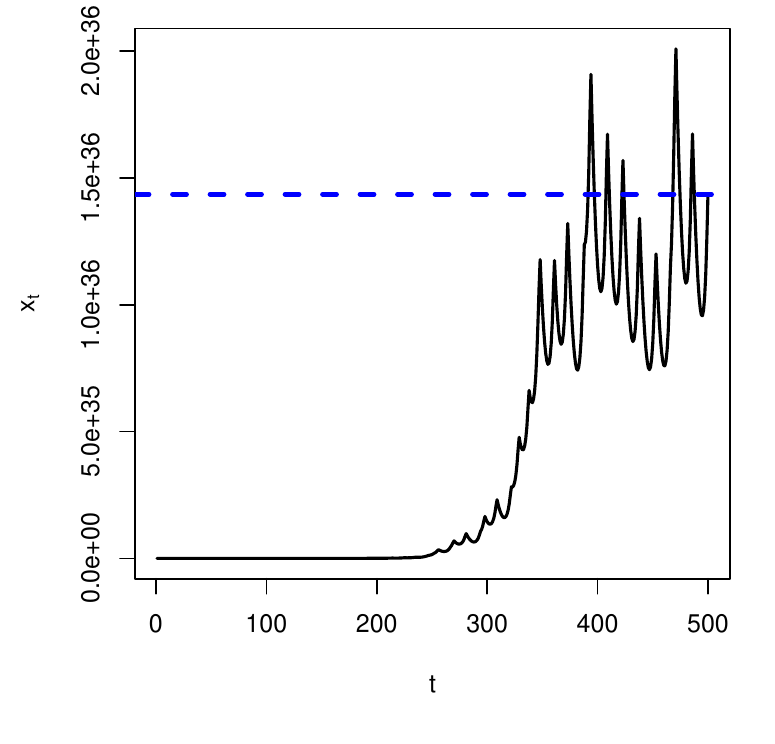} & \includegraphics[scale=0.6]{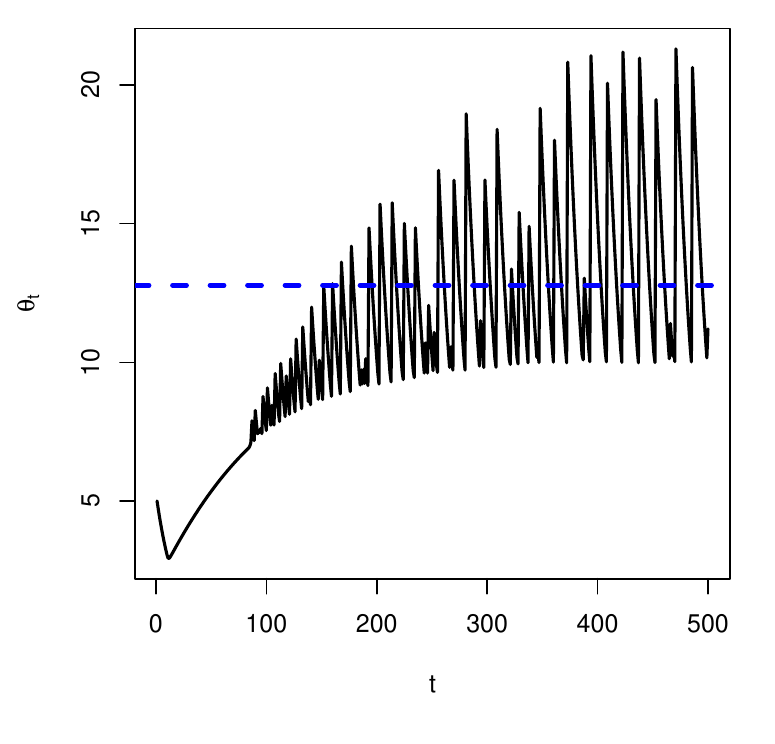} 
   \end{tabular}
   \caption{Times series of a Darwinian model  when $\sigma$ is large. We observe that there are oscillations making the critical point unstable  (blue dashed line).}
   \label{fig:FixPointsOscill}
\end{figure}
 \subsection*{Discussion}

 Assumption $A_1$ is important in that we  only require that $G$ be nonnegative and $G\in L^1(\Omega)$, which guarantees that it can be transformed into the density of a random variable. It does not however guarantees that we can easily obtain a sample from it! If $G$ happens to be a classical distribution (normal, exponential, t-distribution, Weibull, etc.), then there  are sampling methods already available. If $G$ is has a non-classic expression, we may have to resort to either the Probability Integral Transform (see Theorem 2.1.10, p.54 in  \cite{Casella2002}) or to Markov Chain Monte Carlo (MCMC) to obtain a sample, which sometime are themselves onerous in time. Obtaining an efficient estimator of $\theta$ is easily done when $G$ is a classic distribution. While efficiency would be great, it may not be necessary since overtime, the estimator would still converge, albeit slowly, to the fixed point of $\theta$. We observe that estimates we obtain in this case are point estimates of $T_n(\theta_t)$ (mean, median, etc.), albeit, at each time $t$. A Bayesian estimate is also possible, provided that the initial distribution of $\theta_0$ be selected from a well-defined Jeffrey's prior. As for answers to the questions raised in the introduction, we can say based on the above that the set of points $(x,\theta)$ where  $G(x,\theta)$ is maximized at the same  where $\ln(G(x,\theta))$ is minimized, and contains all the critical points of the function $G$. This can be written formally as  $\mbox{Arg}\max_{\theta \in \Omega} G(x,\theta)=\mbox{Arg}\min_{\theta \in \Omega} \ln(G(x,\theta))\supseteq \set{(x,\theta) \in \mathcal{X}\times \Omega: g(x,\theta)=0}$.

\subsection{Single population model with multiple traits}

Now suppose we are in the presence of one species with density $x$ possessing $n$ traits given by the vector $\Theta=(\theta_1, \theta_2, \cdots, \theta_n)$ and a  vector ${\bf U}=(u_1, u_2, \cdots, u_n)$.\\
$H_1$) We assume that $b(\Theta)=b_0\mbox{exp}\left(\ds -\sum_{i=1}^n \frac{\theta_i^2}{2w_i^2}\right)$ is the joint distribution of the independent traits $\theta_i$, each  with mean $0$ and variance $w_i^2$. \\
$H_2$) We also assume that $c_{{\bm U}}(\Theta)=c_0\mbox{exp}\left(\ds -\sum_{i=1}^n \kappa_i(\theta_i-u_i)\right)$\;. \\
$H_3$) We assume that  density of $x_t$ is given as $G(x, \Theta,\bm{U})=b(\Theta)\mbox{exp}(-c_{{\bm U}}(\Theta)x)$ at $t=1$\;. \\
Under $H_1, H_2$, and $H_3$,  we will consider the discrete dynamical system 
\begin{equation} \label{eqn3}
\begin{cases}
x_{t+1}&\vspace{0.5cm}=x_tG(x_t,\Theta_t,U_t)\\
\Theta_{t+1}&=\Theta_t+\Sigma g(x_t, \Theta_t, U_t)\;,
\end{cases}
\end{equation}
where 
\begin{equation}\label{eqn4}
\Sigma=\begin{pmatrix} 
\sigma_{11} & \sigma_{12} &\cdots & \sigma_{1n}\\
\sigma_{21} & \sigma_{22} &\cdots & \sigma_{2n}\\
\vdots & \vdots& \cdots & \vdots\\
\sigma_{n1} & \sigma_{n2} &\cdots & \sigma_{nn}
\end{pmatrix}\;.
\end{equation}
\begin{remark} There are couple of distinctions between this model and the ones encountered in the recent literature,  see for instance \cite{Cushing2019,Elaydi2022}. Firstly,  the matrix $\Sigma$ considered here  is not necessarily symmetric ($\sigma_{ij}\neq \sigma_{ji}$). Secondly,  the competition function $c_{\bm{U}}(\Theta)$, depends subtly on  a vector ${\bf U}$ that needs not be the mean of $\Theta$ as it is often considered. We note however that when $\Theta={\bm U}$, then $c_{{\bm U}}(\Theta)=c_0$. Ecologically, this happens when competition is maximal. This leads to recover the uncoupled Darwinian model (2) in \cite{Elaydi2022}.
\end{remark}
\noindent The result below shows how to obtain the Fisher's information of the Darwinian dynamical system  \eqref{eqn3}.
\begin{thm}\label{thm:SingleSpecMultTraits}
Let $\ds\Gamma_{\bm{U}}(\Theta)=\frac{b(\Theta)}{c_{\bm{U}}(\Theta)}$. Then under assumptions $H_1, H_2$, and $H_3$,  the dynamical system above has  Fisher's information matrix $I(\Theta)$ is given as 
\begin{equation}\label{eqnFishInf}
I(\Theta)=\begin{pmatrix} 
\frac{1}{w_1^2}+\kappa_1^2 \Gamma_{\bm{U}}(\Theta)& \kappa_1\kappa_2 \Gamma_{\bm{U}}(\Theta)&\cdots &  \kappa_1\kappa_n\Gamma_{\bm{U}}(\Theta)\\
 \kappa_1\kappa_2 \Gamma_{\bm{U}}(\Theta) & \frac{1}{w_2^2}+\kappa_2^2\Gamma_{\bm{U}}(\Theta) &\cdots &  \kappa_2\kappa_n\Gamma_{\bm{U}}(\Theta)\\
\vdots & \vdots& \cdots & \vdots\\
 \kappa_1\kappa_n \Gamma_{\bm{U}}(\Theta)&  \kappa_2\kappa_n\Gamma_{\bm{U}}(\Theta)&\cdots &\frac{1}{w_n^2}+\kappa_n^2\Gamma_{\bm{U}}(\Theta)
\end{pmatrix}\;.
\end{equation}
\end{thm}
\noindent The Proof of this result can be found in Appendix B.\\

\noindent A necessary condition for the existence of an extinction equilibria  $(0, {\bm 0})$ for this this system is that $\mbox{det}(\Sigma)\neq 0$. Letting  $\rho(A)$ represent the spectral radius of matrix $A$,  it was  proved  in \cite{Cushing2019}  that if $\rho({\bm I}+\Sigma h(0,{\bm 0}))<1$ and $b_0<1$ , then the extinction equilibria is asymptotically stable and unstable if $\rho({\bm I}+\Sigma h(0,{\bm 0}))<1$ and $b_0>1$,   or if $\rho({\bm I}+\Sigma h(0,{\bm 0}))>1$ for all $b_0>0$. This is particularly true if  $\Sigma$ is diagonally dominant and for  $\sigma_{12}=\sigma_{21}=0$. This  system admits  a nontrivial fixed point (positive equilibria)  $ (x_*,\Theta_{*})$ if   $G (x_*,\Theta_*)=1$ and $\mbox{det}(\Sigma)= 0$. In the proposition below, we give a more precise characterization of nontrivial fixed points of the system \eqref{eqn:syst}. 

\begin{proposition}\label{propFixPointsGen}
Assume $\mbox{det}(\Sigma)= 0$ and $\sigma_{ii}\neq 0$  for $i,j=1, \cdots, n$.  Put 
\[\ds \quad \mu_{ij}=\frac{\sigma_{ij}}{\sigma_{ii}},\quad \nu_j=\kappa_j+\sum_{i=1,i\neq j}^n \mu_{ij}\kappa_i\;,\] 
and given $j\in \set{1,2,\cdots,n}$
\[\xi_{nj}:=\frac{1}{\nu_j^2}\left(\frac{1}{w_j^2}+\sum_{i=1,i\neq j}^n\frac{\mu_{ij}^2}{w_i^2}\right)+2\ln(b_0)\;.
\]
If  $\xi_{nj}<0$, then there is nonon trivial solution for the system \eqref{eqn3}.\\
Now suppose $\xi_{nj} \geq 0$. Then the system  \eqref{eqn3} has a nontrivial solution $ (x_*,\Theta_*)$ given as 
\begin{equation}\label{eqn:xstar}
x_*=\frac{\ds \ln(b_0)-\sum_{i=1}^n \frac{(\theta_{i*})^2}{2w_i^2}}{c_0\mbox{exp}\left(-\ds \sum_{i=1}^n\kappa_i(\theta_{i*}-u_i)\right)}\;,
\end{equation}
and 
\begin{itemize}
\item[(i)] if $\xi_{nj}=0$, then $\Theta_*=\left(-\frac{1}{\nu}, -\frac{\mu_2}{\nu}, \cdots, -\frac{\mu_n}{\nu}\right)$,
\item[(ii)]  if $\xi_{nj}>0$, then  the coordinates of the vector $\Theta_*$ are   points  that lie on  the curve of equation 

\begin{equation}\label{eqnEllipseGen}
\frac{1}{w_j^2}\left(\theta_j+\frac{1}{\nu}\right)^2 +\sum_{i=1,i\neq j}^n \frac{1}{w_i^2}\left(\theta_i+\frac{\mu_{ij}}{\nu_j}\right)^2-\xi_{nj}=0\;.
  \end{equation}

\end{itemize}

\end{proposition}
\noindent The proof can be found in Appendix C.\\

\noindent {\bf Special case: single species with  two traits}\mbox{}\\

\noindent Here we consider the  particular case of system of   one  species with  two traits, namely the coupled  dynamical system 
\begin{equation} \label{eqn55}
\begin{cases}
x_{t+1}&=x_tG(x_t,\theta_{1,t},\theta_{2,t})\\
\theta_{1,t+1}&=\theta_{1,t}+\sigma_{11} g_1(x_t, \theta_{1,t},\theta_{2,t})+\sigma_{12} g_2(x_t, \theta_{1,t},\theta_{2,t})\\
\theta_{2,t+1}&=\theta_{2,t}+\sigma_{21} g_1(x_t, \theta_{1,t},\theta_{2,t})+\sigma_{22} g_2(x_t, \theta_{1,t},\theta_{2,t})\\
\end{cases}\;.
\end{equation}
The   fixed points $(x_*, \theta_{1*},\theta_{2*})$ of this model are solutions of  the system of equations
\begin{equation}\label{eqn:syst}
\begin{cases}
xG(x,u,v)&=x\\
\Sigma g(x,u,v)&=0
\end{cases}\;,
\end{equation}
where 
\[
\Sigma=\begin{pmatrix}
\sigma_{11}&  \sigma_{12}\\
\sigma_{21} & \sigma_{22} 
 \end{pmatrix}\;, \quad \mbox{for $(u,v)\in \R^2$}\;.
\]
\begin{corollary}\label{propFixPoints}
Assume $\sigma_{11},\sigma_{22}\neq 0$ and $\sigma_{11}\sigma_{22}=\sigma_{12}\sigma_{21}$.  We  put 
\[\ds \mu_1=\frac{\sigma_{21}}{\sigma_{22}}, \quad \mu_2=\frac{\sigma_{12}}{\sigma_{11}},\quad \nu=\kappa_1+\mu_2\kappa_2\;,\] 
and 
\[\xi_2:=\frac{1}{\nu^2}\left(\frac{1}{w_1^2}+\frac{\mu_2^2}{w_2^2}\right)+2\ln(b_0)\;.
\]
If  $\xi_2<0$, then there is no non trivial solution for the system \eqref{eqn55}.\\
Now suppose $\xi_2 \geq 0$. Then the system \eqref{eqn55} has a non trivial solution $ (x_*,\theta_{1*},\theta_{2*})$ given as 
\begin{equation}\label{eqn:xstar}
x_*=\frac{\ds \ln(b_0)-\sum_{i=1}^n \frac{(\theta_{i*})^2}{2w_i^2}}{c_0\mbox{exp}\left(-\ds \sum_{i=1}^n\kappa_i(\theta_{i*}-u_i)\right)},\quad \mbox{for $n=2$}\;.
\end{equation}
and 
\begin{itemize}
\item[(i)] if $\xi_2=0$, then $(\theta_{1*},\theta_{2*})=\left(-\frac{1}{\nu}, -\frac{\mu_2}{\nu}\right)$,
\item[(ii)]  if $\xi_2>0$, then  $(\theta_{1*},\theta_{2*})$ are   points  that lie on  the ellipse of equation 

\begin{equation}\label{eqnEllipse}
\frac{\left(\theta_1+\frac{1}{\nu}\right)^2}{a^2}+ \frac{\left(\theta_2+\frac{\mu_2}{\nu}\right)^2}{b^2}=1, \quad \mbox{where $a=w_1\sqrt{\xi_2}$  and $b=w_2\sqrt{\xi_2}$}\;.
  \end{equation}

\end{itemize}

\end{corollary}

\noindent In Figure \ref{fig:fixPoitnts} below, we illustrate Proposition \ref{propFixPoints} for $x_0=1,\theta_{10}=3; \theta_{20}=5; w_1=4; w_2=1; \kappa_1=3; \kappa_2=0.5,\sigma{11}=0.1; \sigma_{21}=1; \sigma_{11}=1; \sigma_{12}=2; \sigma_{22}=2; u_1=3; u_2=0, c_0=0.1$. We verify that $\mu_1=\sigma_{21}/\sigma_{22}=0.5$ and $\mu_2=\sigma_{12}/\sigma_{11}=2$. Therefore,  $\mu_1\mu_2=1$, that is, $\mbox{det}(\Sigma)=0$. We also have that $\nu=4$ and that $\xi_2\approx 4.86>0$. Hence according to proposition \ref{propFixPoints} above, $(\theta_{1*},\theta_{2*})$ is expected to be on the ellipse centered as $(-1/\nu,\mu_2/\mu)=(-0.25,-0.5)$ with respective major and minor axis lengths $a=w_1\sqrt{\xi}\approx 8.82$ and $b=w_2\sqrt{\xi}\approx 2.21$.\\
\noindent The parameters for Figure \ref{fig:fixPoitntsStochast} are the same except for $\theta_{10}=3, \theta_{20}=1$ and $x_0$ is generated from an exponential distribution with parameters $c_{{\bm U}}(\Theta)=c_0\mbox{exp}\left(-\ds \sum_{i=1}^n\kappa_i(\theta_{i0}-u_i)\right)$.
\begin{figure}[H] 
   \centering
   \begin{tabular}{cc}
   {\bf (a)} & {\bf (b)}\\
      \includegraphics[scale=0.58]{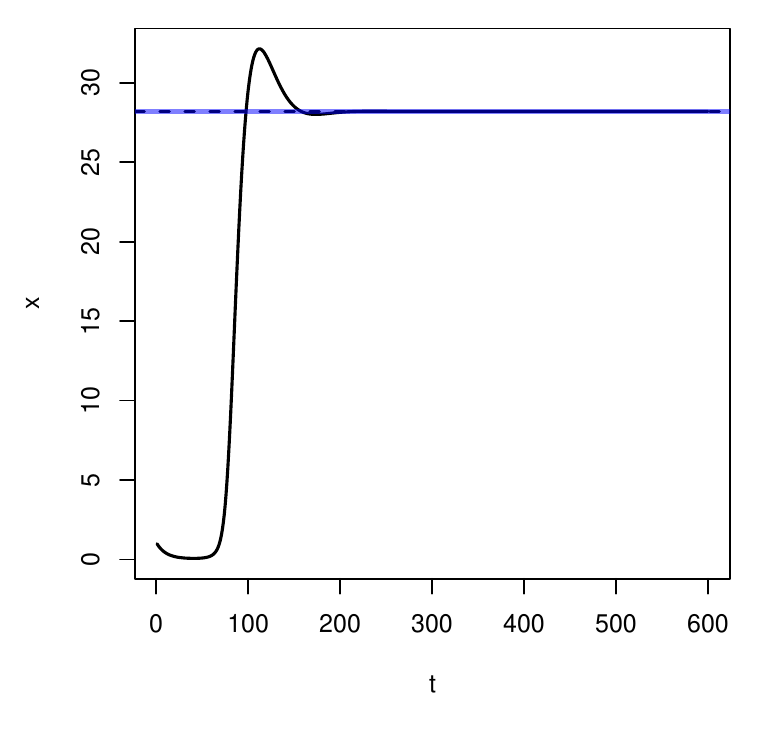} & \includegraphics[scale=0.58]{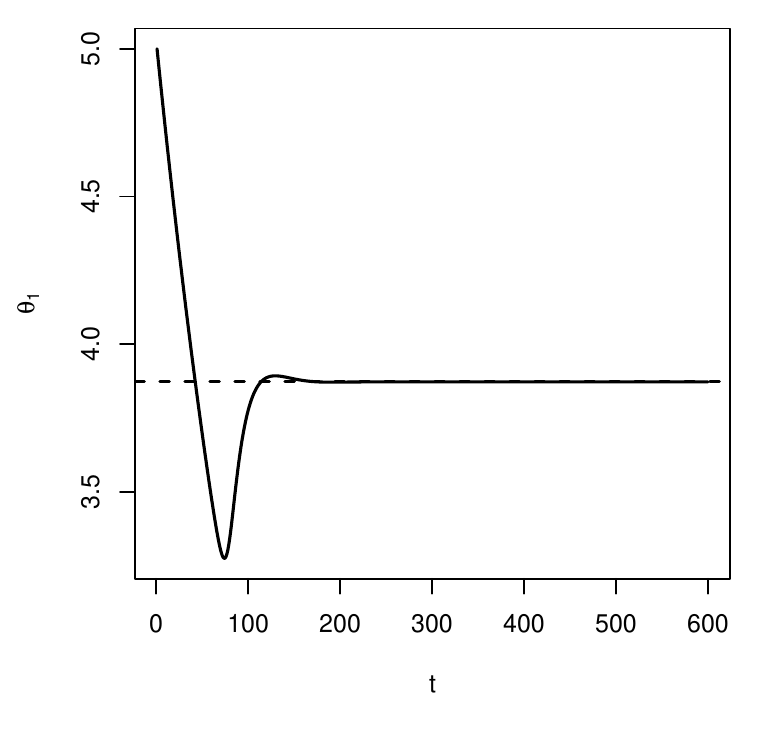}\\
      {\bf (c)} & {\bf (d)}\\
        \includegraphics[scale=0.58]{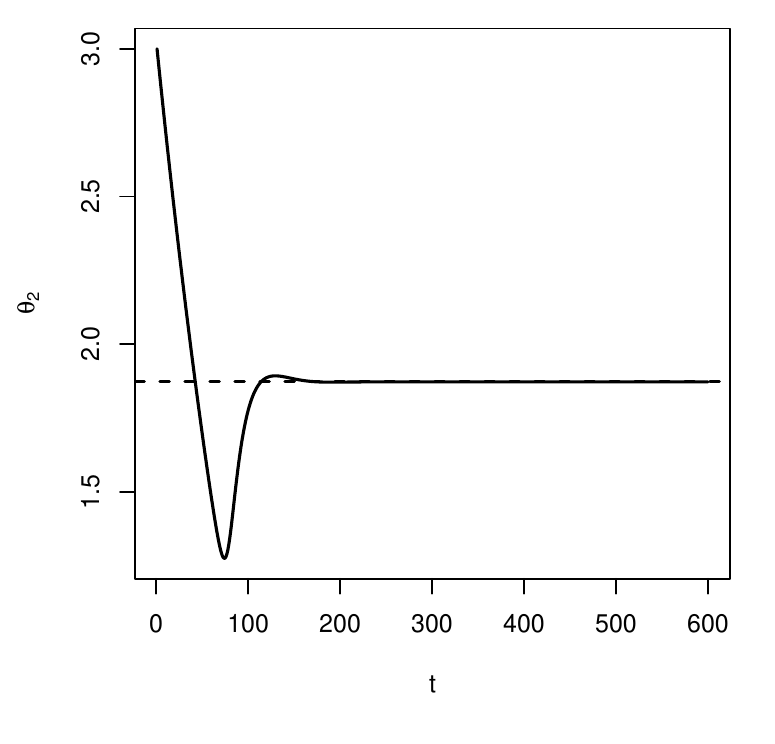} & \includegraphics[scale=0.58]{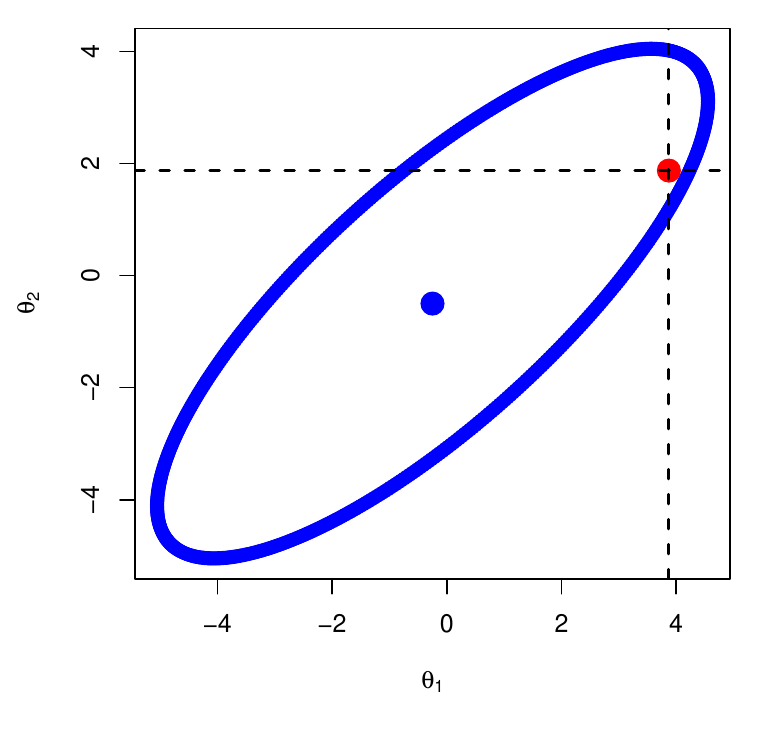}
   \end{tabular}
   \caption{In {\bf (a)}, the solid curve represents the dynamics of $x_t$ in the system \eqref{eqn55} above. The dashed line represents the nontrivial equilibrium point $x_*$. The blue line represent the value of $x_*$ as given in equation \eqref{eqn:xstar}, using the values of $\theta_{1*}\approx 3.873$ and $\theta_{2*}\approx 1.873$ obtained as nontrivial fixed points from the last two equations in \eqref{eqn55}. That the blue line and the dashed coincide is a proof of the first part of the Proposition above. In {\bf (b)} and {\bf (c)}, the solid curves represents the dynamics of $\theta_{1,t}$ and $\theta_{2,t}$ respectively. The dashed lines line represent the nontrivial fixed points $(\theta_{1*}, \theta_{2*})$. In {\bf (d)}, the blue curve represents the ellipse given in equation \eqref{eqnEllipse} above, with center $(-1/\nu, -\mu_2/\nu)$. The red dot represents the nontrivial fixed  $(\theta_{1*}, \theta_{2*})$. This point almost lies on the ellipse (the discrepancy is due to an accumulation of error) is a proof of the second part of   Proposition \ref{propFixPointsGen} above. }
   \label{fig:fixPoitnts}
\end{figure}
\begin{figure}[htbp] 
   \centering
   \begin{tabular}{cc}
   {\bf (a)} & {\bf (b)}\\
      \includegraphics[scale=0.6]{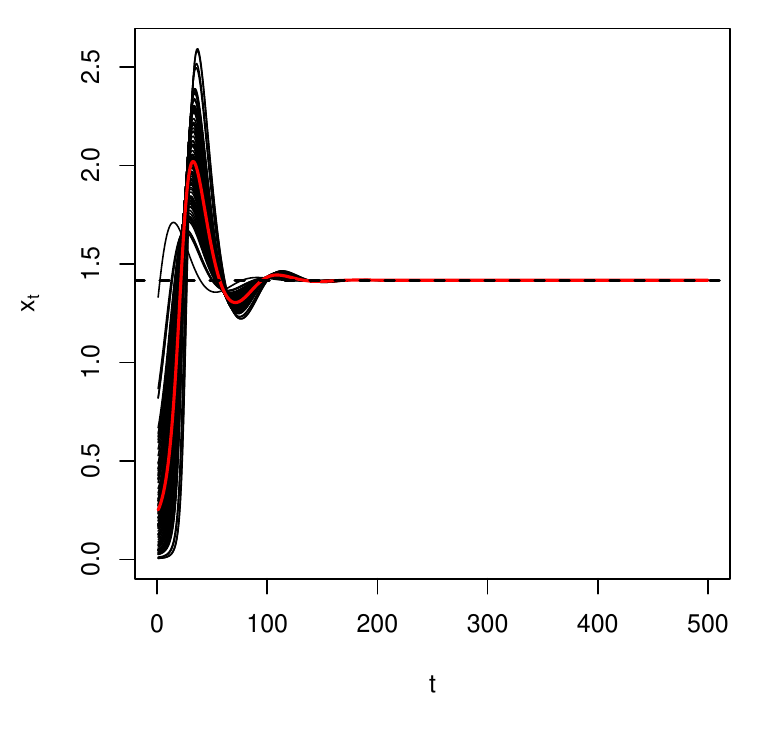} & \includegraphics[scale=0.6]{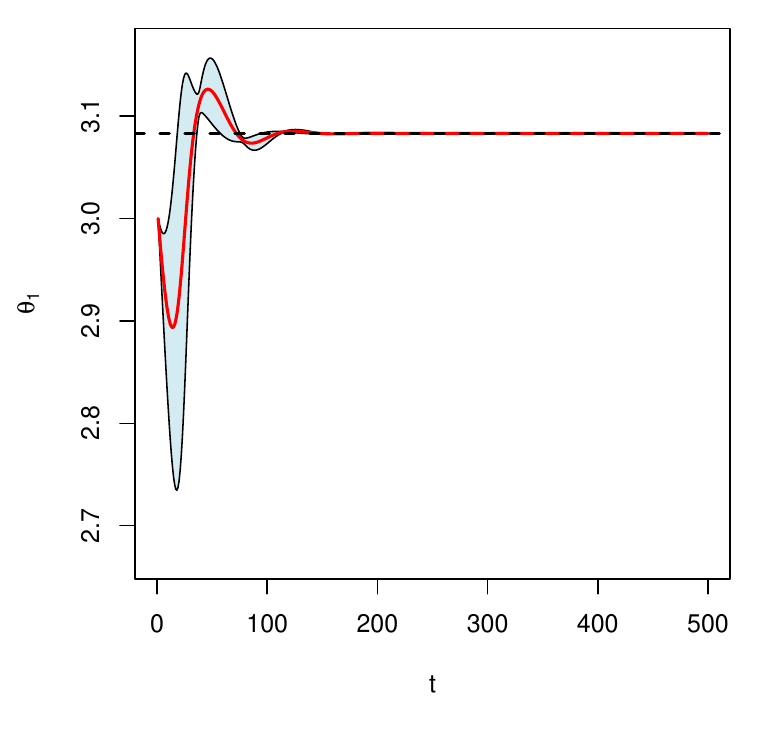}\\
      {\bf (c)} & {\bf (d)}\\
        \includegraphics[scale=0.6]{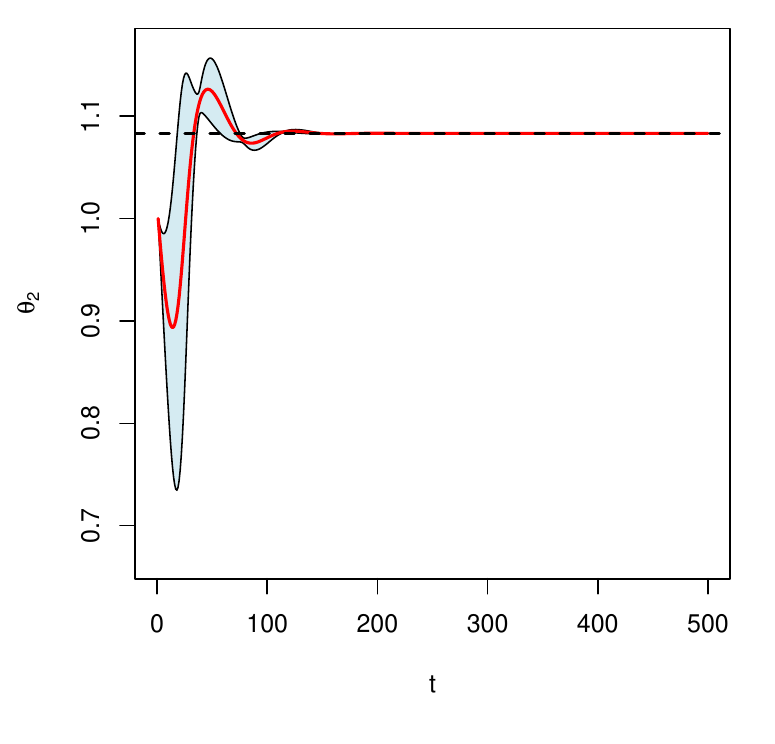} & \includegraphics[scale=0.6]{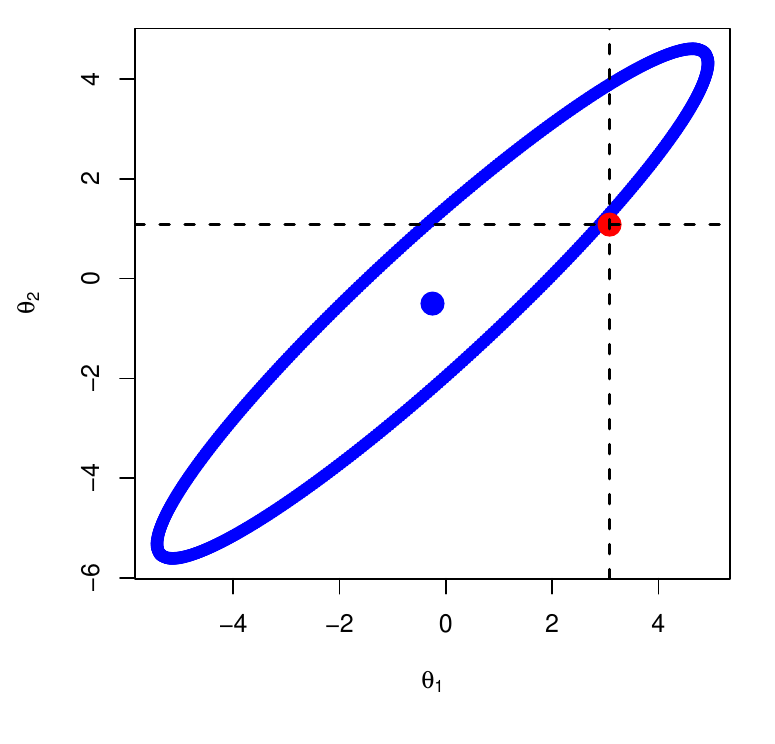}
   \end{tabular}
   \caption{In {\bf (a)}, represented in black are 100 trajectories of $x_t$ with a starting point selected at random from an exponential distribution with parameter $c_u(\theta)$. The red  curve represents their average over time converging to $x_*\approx 1.417$. In {\bf (b)}, represented in light-blue are the  95\% confidence bands for the corresponding trajectories of $\theta_{1,t}$. The red curve represents their average and we verify that they all converge to $\theta_{1*}\approx 3.083$. We note that these confidence bands are constructed using the Fisher's information as $\overline{\theta_{2,t}}\pm 1.96/\sqrt{I_{11}(\Theta_t)}$, where $\overline{\theta_{1,t}}$ is the average at time $t$. Similarly in {\bf (c)}, represented in light-blue are the 95\% confidence bands for $\theta_{2,t}$ and the corresponding sample average in red. They converge to $\theta_{2*}\approx 1.083$.  In { \bf (d)}, represented is the ellipse given in equation \eqref{eqnEllipse} above. We verify that the point $(\theta_{1*}, \theta_{2*})$ is on the ellipse and that the value of $x^*$ obtained from equation \eqref{eqn:xstar} is the same as  convergence value of $x_t$.}
   \label{fig:fixPoitntsStochast}
\end{figure}
\subsection*{Discussion}

\noindent An important remark about Theorem \ref{thm:SingleSpecMultTraits} is that we assume the vector $\bm{U}$ is given. In fact, if the vector $\Theta$ were given, the same technique could have been used for the estimation of $\bm{U}$, up to a negative sign on the Fisher's information matrix. Having $\bm{U}$ be different from the average of $\Theta$ allows for generalization in that,  $\Theta-\bm{U}$ represents the difference of a set of traits $\Theta$ from a given set  of traits $\bm{U}$, which needs not be the average of $\Theta$. One thing we have not insisted much in this paper is the type of  estimator of $\Theta$ itself. 
We do not need to specify in particular which estimator to use, since the inverse of the Fisher's information is smallest variance for all estimators.
 One consequence of  Proposition \ref{propFixPointsGen} above is that  once we have estimated $\Theta_*$, we can deduct the value of $x_*$. The results above also show that there may be many equilibria when $\xi_n>0$. In the context of evolution and natural selection, one should focus  on equilibria that ensure better adaptation to environmental fluctuations. This specifically means adding stochasticity to the model by means of, say  a Wiener process and finding  traits that ensure for example that on average, the species density is bounded away from the extinction equilibrium.  Another possibility could be  to focus on equilibria  that maximize the species density in order to increase the prospects of survivability of the species. This can be done using a constraint optimization problem $\underset{{\bm u}\in \R^n} \max~f({\bm u})$ where $f({\bm u})=x_*$ subject to the constraint  that the point ${\bm u}$ be on the curve   defined in equation \eqref{eqnEllipseGen}.  
In the particular case where  $n=2$ with ${\bm u}=(u,v)$, we have 
\[
f(u,v):=\frac{\ds \ln(b_0)- \frac{u^2}{2w_1^2}-\frac{v^2}{2w_2^2}}{c_0\mbox{exp}\left(-\kappa_1(u-u_i)-\kappa_2(v-u_i)\right)}\;.
\]
However, the form of the function $f(u,v)$ makes it a very challenging problem.  Rewriting, we have 
\[
f(u,v):=\ds c_0^{-1} \left(\ln(b_0)- \frac{u^2}{2w_1^2}-\frac{v^2}{2w_2^2}\right)e^{\kappa_1(u-u_i)+\kappa_2(v-u_i)}\;.
\]
We observe that for positive $\kappa_1$ and $\kappa_2$,  the function $(u,v)\mapsto e^{\kappa_1(u-u_i)+\kappa_2(v-u_i)}$ is a positive and  increasing function of $u$ and $v$. Therefore, maximizing $f(u,v)$ amounts to maximizing 
\begin{equation} \label{eqn:fstar}
f_*(u,v)=\ln(b_0)- \frac{u^2}{2w_1^2}-\frac{v^2}{2w_2^2}\;.
\end{equation}
Geometrically, this is the equation of a paraboloid that bends down. Therefore, the constraint optimization amounts to  finding the points of intersection between a paraboloid and an ellipse. This means there can be between 0 and 4 points of intersection. Finally,  let us observe that the expression of $x^*$ in the case of multiple traits is just a generalization of the case of one trait. In fact, the first equation in the system \eqref{eqn:OneTrait} can be written as $\ds x^*=\frac{\ln(b(\theta))}{c_u(\theta)}$. We see that this is similar to the expression of $x^*$ given is equation \eqref{eqn:xstar} for two traits, which naturally generalizes to the case of one species with $n \geq 2$ traits. 

\section{Concluding remarks}\label{sect5}

In this paper, we have proposed a way of estimating traits coefficients in  a Darwinian evolution population dynamics model using a Fisher's information matrix. In particular, we proposed  how to characterize uncertainty on the estimation process. We have discussed specifically  the cases of one species with one trait and one species with multiple traits. The case of multiple species with one or more traits is certainly an interesting one to tackle and is left as future research . We have proposed a relationship between a G-function, its natural logarithm, and  its derivative $g$.
A bi-product of our work provide a precise characterization of  nontrivial fixed points of the model. We showed that once the critical density $x^*$ has been found, the set of critical traits $\Theta_*$ lie on a well-defined curve in $\R^n$. On the other hand, if the set critical traits $\Theta_*$ is given, there may be a unique critical density $x_*$ for the Darwinian system which is not necessarily  appropriate for the survival of the species. Another approach to the estimation of traits parameters would be to use modern machine learning techniques. Indeed, traits parameters  may  be estimated by minimizing the relative information or Kullback-Leibler divergence  in  a Darwinian evolution population model using either a classical gradient ascent or a stochastic gradient ascent. This may require identification of appropriate weights for the minimization process. In both cases, the procedure has to  be done talking into account    supervised or unsupervised learning environments. An extension of this work would be to add stochasticity in the form of Wiener process to the model in order to study   strong persistence on average, and 
the existence of global solutions and   stationary distributions. \bibliography{EVP}

\begin{thebibliography}{22}
\providecommand{\natexlab}[1]{#1}
\providecommand{\url}[1]{\texttt{#1}}
\expandafter\ifx\csname urlstyle\endcsname\relax
  \providecommand{\doi}[1]{doi: #1}\else
  \providecommand{\doi}{doi: \begingroup \urlstyle{rm}\Url}\fi

\bibitem[Abbott and Dayan(1999)]{Abbott1999}
L.~F. Abbott and P.~Dayan.
\newblock The effect of correlated variability on the accuracy of a population
  code.
\newblock \emph{Neural Computations}, 11\penalty0 (1):\penalty0 91--101, 1999.

\bibitem[Ackleh et~al.(2015)Ackleh, Cushing, and Salceneau]{Ackleh2015}
A.~S. Ackleh, J.~M. Cushing, and P.~L. Salceneau.
\newblock On the dynamics of evolutionary competition models.
\newblock \emph{Natural Resource Modelling}, 28\penalty0 (4):\penalty0
  380--397, 2015.

\bibitem[Bernado and Smith(1994)]{Bernado1994}
J.~M. Bernado and A.~F.~M Smith.
\newblock \emph{Bayesisan Theory}.
\newblock John Wiley and Sons, 1994.

\bibitem[Casella and Berger(2002)]{Casella2002}
G.~Casella and R.~L. Berger.
\newblock \emph{Statistical Inference}.
\newblock Cengage, 2nd edition, 2002.

\bibitem[Cram\'er(1946)]{Cramer1946}
H.~Cram\'er.
\newblock \emph{Mathematical Methods of Statistics}.
\newblock Princeton Univ. Press, 1946.

\bibitem[Cushing(2019)]{Cushing2019}
J.~M. Cushing.
\newblock Difference equations as models of evolutionary population dynamics.
\newblock \emph{Journal of Biological Dynamics}, 13\penalty0 (1):\penalty0
  103--127, 2019.

\bibitem[Cushing et~al.(2023)Cushing, Park, Farrell, and Chitnis]{Cushing2023}
J.~M. Cushing, J.~Park, A~Farrell, and N.~Chitnis.
\newblock Treatment of outocme in an si model with evolutionary resistance: a
  darwinian model for the evolutionary resistance.
\newblock \emph{Journal of Biological Dynamics}, 17\penalty0 (1):\penalty0
  2255061, 2023.

\bibitem[da~Fonseca and Samengo(2016)]{Fonseca2016}
M.~da~Fonseca and I.~Samengo.
\newblock Derivation of human chromatic discrimination ability from an
  information-theoretical notion of distance in color space.
\newblock \emph{Neural Computations}, 28\penalty0 (12):\penalty0 2628--2655,
  2016.

\bibitem[Darwin(1859)]{Darwin1859}
C.~Darwin.
\newblock \emph{On the origin of species by means of natural selection}.
\newblock John Murray, Albemarble Street, London, 1859.

\bibitem[Elaydi et~al.(2022)Elaydi, Kang, and Luis]{Elaydi2022}
S.~Elaydi, Y.~Kang, and R.~Luis.
\newblock The effects of evolution on the stability of competing species.
\newblock \emph{Journal of Biological Dynamics}, 16\penalty0 (1):\penalty0
  816--839, 2022.

\bibitem[Fisher(1922)]{Fisher1922}
R.~A. Fisher.
\newblock On the mathematical foundation of theoretical statistics.
\newblock \emph{Philosophical Transactions of the Royal Society of London Serie
  A}, 222:\penalty0 594--604, 1922.

\bibitem[Kilrkpatrick et~al.(2017)Kilrkpatrick, Pascanu, Rabinowitz, Veness,
  Desjardins, Rusu, Milan, Quand, and Ramalho]{Kilrkpatrick2017}
J.~Kilrkpatrick, R.~Pascanu, N.~Rabinowitz, J.~Veness, G.~Desjardins, A.~A.
  Rusu, K.~Milan, J.~Quand, and T.~Ramalho.
\newblock Overcoming catastrophic forgetting in neural networks.
\newblock \emph{Proceedings of the National Academy of Sciences}, 114\penalty0
  (13):\penalty0 3521--3526, 2017.

\bibitem[Lehman and Casella(1998)]{Lehman1998}
E.~L. Lehman and G.~Casella.
\newblock \emph{Theory of Point Estimation}.
\newblock Springer, 2nd edition, 1998.

\bibitem[Mokni et~al.(2020)Mokni, Elaydi, and Eladdadi]{Mokni2020}
K.~Mokni, S.~Elaydi, and A.~Eladdadi.
\newblock Disctre evolutionary population models: a new approach.
\newblock \emph{Journal of Biological Dynamics}, 14\penalty0 (1):\penalty0
  454--478, 2020.

\bibitem[Parag et~al.(2022)Parag, Donnelly, and Zarebski]{Parag2022}
K.V. Parag, C.~A. Donnelly, and A.~E. Zarebski.
\newblock Quantifying the information in noisy epidemic curves.
\newblock \emph{Nature Computational Science}, 2\penalty0 (9):\penalty0
  584--594, 2022.

\bibitem[Rao(1945)]{Rao1945}
C.~Rao.
\newblock Information and the accuracy attainable in the estimation of
  statistical parameters.
\newblock \emph{Bulletin of the Calcutta Mathematical Society.}, 37:\penalty0
  81--89, 1945.
\newblock Calcutta Mathematical Society.

\bibitem[Smith(1918)]{Smith1918}
K.~Smith.
\newblock On the standard deviations of adjusted and interpolated values of an
  observed polynomial function and its constants and the guidance they give
  towards a proper choice of the distribution of observations.
\newblock \emph{Biometrika}, 1:\penalty0 1--85, 1918.

\bibitem[Smith(1982)]{Smith1982}
M.~Smith.
\newblock \emph{Evolution and the Theory of Games}.
\newblock Cambridge Unievrsity Press, Cambridge, 1982.

\bibitem[Smith and Price(1973)]{Smith1973}
M.~Smith and G.~R. Price.
\newblock The logic of animal conflict.
\newblock \emph{Nature}, 246:\penalty0 15--18, 1973.

\bibitem[Vincent and Brown(2005)]{Vincent2005}
T.~L. Vincent and J.~S. Brown.
\newblock \emph{Evoluationary Game Theory, Natural Selection, and Darwinian
  Dynamics}.
\newblock Cambridge University Press, Cambridge, 2005.

\bibitem[Vincent et~al.(2011)Vincent, Vincent, and Cohen]{Vincent2011}
T.~L. Vincent, T.~L.~S. Vincent, and Y.~Cohen.
\newblock Darwinian dynamics and evolutionary game theory.
\newblock \emph{Journal of Biological Dynamics}, 5\penalty0 (3):\penalty0
  215--226, 2011.

\bibitem[Ward and Ahlquist(2015)]{Ward2015}
M.~D. Ward and J.~Ahlquist.
\newblock \emph{Maximum Likelihood for Social Science: Strategies for
  Analysis}.
\newblock Cambridge University Press, 2015.
\newblock ISBN ISBN 978-1-316-63682-4.

\end{thebibliography}
\section{Appendix}


\subsection{Appendix $A_1$: Proof of Theorem \ref{thm:FishOneSpecOneTrait}}

\begin{proof}
We have that $\ds \lambda(x,\theta)=\ln(G(x,\theta))=\ln(b_0)-\frac{\theta^2}{2w^2}-c_0xe^{-\kappa(\theta-u)}$.\\
It follows that \begin{eqnarray*}
\ds g(x,\theta)&=&-\frac{\theta}{w^2}+\kappa c_0x e^{-\kappa(\theta-u)}\\
&=& -\frac{\theta}{w^2}+\kappa c_u(\theta)x\;, 
\end{eqnarray*} and \[h(x,\theta)=-\frac{1}{w^2}-\kappa^2 c_0 x e^{-\kappa(\theta-u)}\;.\] From Definition \ref{def1} above, it follows that 
\[I(\theta)=-\mathbb{E}_X[h(X,\theta)]=-\mathbb{E}_X\left[-\frac{1}{w^2}-\kappa^2c_u(\theta)X\right]=\frac{1}{w^2}+\kappa^2c_u(\theta)\mathbb{E}[X]\;.\]
We observe that $X$ has probability distribution $G(x,\theta)=b(\theta)e^{-c_u(\theta)x}$, therefore, 
\begin{eqnarray*}
\mathbb{E}[X]&=& \int_{\mathcal{X}} xG(x,\theta)dx\\
&=& \frac{b(\theta)}{c_u(\theta)}\int_{\mathcal{X}} xc_u(\theta)e^{-c_u(\theta)x}dx\\
&=& \frac{b(\theta)}{c_u(\theta)^2}\;.\\
\end{eqnarray*}
  It then follows  that 
  \begin{equation}\label{eqn:ProofFishOneTrait}
  I(\theta)=\frac{1}{w^2}+\kappa^2 c_u(\theta) \frac{b(\theta)}{c_u(\theta)^2}=\frac{1}{w^2}+\kappa^2 \frac{b(\theta)}{c_u(\theta)}=\frac{1}{w^2}+\kappa^2 \Gamma(\theta)\;.
  \end{equation}
  This ends the proof of the Theorem. 
\end{proof}
\subsection{Appendix $A_2$: Proof of Corollary \ref{cor:FishOneSpecOneTrait}}
Suppose $\ds \frac{b_0}{c_0}= \frac{1}{\kappa\sqrt{2\pi w \kappa}}e^{\kappa u-\frac{(w\kappa)^2}{2}} $.
From equation \eqref{eqn:ProofFishOneTrait} above, we have that 
\[I(\theta)=\frac{1}{w^2}+\kappa^2 \frac{b_0}{c_0}e^{-\frac{\theta^2}{2w^2}+\kappa(\theta-u)}\;.\]
Completing the square, we have that 
\begin{eqnarray*}
-\frac{\theta^2}{2w^2}+\kappa(\theta-u)&=& -\frac{1}{2w^2}\left[\theta^2-2w^2\kappa \theta+2w^2\kappa u\right]\\
&=& -\frac{1}{2w^2}\left[ (\theta-w^2\kappa)^2-w^4\kappa^2+2w^2\kappa u\right]\\
&=& -\frac{1}{2w^2}\left[ (\theta-w^2\kappa)^2\right]+\frac{1}{2}(w\kappa)^2-\kappa u\;.
\end{eqnarray*}
It follows that 
\begin{eqnarray*}
I(\theta)&=& \frac{1}{w^2}+\kappa^2 \frac{b_0}{c_0}e^{\frac{1}{2}(w\kappa)^2-\kappa u} e^{-\frac{1}{2w^2}\left[ (\theta-w^2\kappa)^2\right]}\;.
\end{eqnarray*}
From the given expression of  $\ds \frac{b_0}{c_0}$ above, it follows that  that  
\[I(\theta)=\frac{1}{w^2}+\frac{1}{w\sqrt{2\pi}}e^{-\frac{1}{2w^2}\left[ (\theta-w^2\kappa)^2\right]}\;. \]
\subsection{Appendix B: Proof of Theorem \ref{thm:SingleSpecMultTraits}}

\begin{proof}
Let $b(\Theta)=b_0\mbox{exp}\left(\ds -\sum_{i=1}^n \frac{\theta_i^2}{2w_i^2}\right)$ and $c_{\bm{U}}(\Theta)=c_0 \mbox{exp}\left(\ds -\sum_{i=1}^n \kappa_i(\theta_i-u_i)\right)$.\\
We have that \begin{eqnarray*}
\ds G(x,\Theta)&=& b_0\mbox{exp}\left(\ds -\sum_{i=1}^n \frac{\theta_i^2}{2w_i^2}-c_0 x \mbox{exp}\left(\ds -\sum_{i=1}^n \kappa_i(\theta_i-u_i)\right)\right)\\
&=& b(\Theta)e^{-c_{\bm{U}}(\Theta)x}\;,
\end{eqnarray*}
 from which we can deduct \begin{eqnarray*}
 \lambda(x,\Theta)&=&\ln(b_0)-\sum_{i=1}^n \frac{\theta_i^2}{2w_i^2}-c_0 x\mbox{exp}\left(\ds -\sum_{i=1}^n \kappa_i(\theta_i-u_i)\right)\\
 &=& \ln(b(\Theta))-xc_{\bm{U}}(\Theta)\;.
 \end{eqnarray*}
 It follows that $g(x,\Theta)$ is the vector given as 
\[g(x,\Theta):=(g_i(x,\Theta))_{i=1,\cdots, n}:=\left(-\frac{\theta_i}{w_i^2}+c_0\kappa_i x\mbox{exp}\left(\ds -\sum_{i=1}^n \kappa_i(\theta_i-u_i)\right)\right)_{i=1, 2, \cdots,n}\;.\]
Hence,  $h(x,\Theta)$ is an $n\times n$ matrix given by 
\[
h(x,\Theta)=\begin{pmatrix}
g_{11}(x,\Theta) & g_{12}(x,\Theta) & \cdots & g_{1n}(x,\Theta) \\
g_{21}x,\Theta)  & g_{22}(x,\Theta) & \cdots & g_{2n}(x,\Theta) \\
\vdots & \vdots& \cdots&  \vdots\\
g_{n1}(x,\Theta)  & g_{n2}(x,\Theta) & \cdots & g_{nn}(x,\Theta) \\
\end{pmatrix}\;,
\]
where for $i=1,2,\cdots, n$, we have  \begin{eqnarray*}
g_{ii}(x,\Theta) =\frac{\partial^2 \lambda}{\partial \theta_i^2}&=&-\frac{1}{w_i^2}-c_0\kappa_i^2 x\mbox{exp}\left(\ds -\sum_{i=1}^n \kappa_i(\theta_i-u_i)\right)=-\frac{1}{w_i^2}-\kappa_i^2 c_{\bm{U}}(\Theta)x\;.
\end{eqnarray*}
 and \begin{eqnarray*}
 g_{ij}(x,\Theta) =\frac{\partial^2 \lambda}{\partial \theta_i \partial \theta_j}&=&-c_0\kappa_i\kappa_j x \mbox{exp}\left(\ds -\sum_{i=1}^n \kappa_i(\theta_i-u_i)\right)=\kappa_i\kappa_j c_{\bm{U}}(\Theta)x\;.
 \end{eqnarray*}
We can deduct that $I(\Theta)$ is the $n\times n$ matrix given as 
\[
\begin{pmatrix}
I_{11}(\Theta) & I_{12}(\Theta) & \cdots &  I_{1n}(\Theta) \\
I_{21}x,\Theta)  & I_{22}(\Theta) & \cdots &  I_{2n}(\Theta) \\
\vdots & \vdots& \cdots & \vdots\\
I_{n1}(\Theta)  & I_{n2}(\Theta) & \cdots&  I_{nn}(\Theta) \\
\end{pmatrix}\;,
\]
where \[I_{ii}(\Theta)=-\E_X\left[\frac{\partial^2 \lambda(X,\Theta)}{\partial \theta_i^2}\right]=\frac{1}{w_i^2}+\kappa_i^2 c_{\bm{U}}(\Theta) \E[X]\;.\] Since $X$ has distribution  $G(x,\Theta)$ 
we have that \begin{eqnarray*}
\E[X]&=& \int_{\mathcal{X}}xG(x,\Theta)dx\\
&=& \frac{b(\Theta)}{c_{\bm{U}}(\Theta)}  \int_{\mathcal{X}} x c_{\bm{U}}(\Theta)e^{-c_{\bm{U}}(\Theta)x}dx\\
&=& \frac{b(\Theta)}{c_{\bm{U}}(\Theta)}\cdot \frac{1}{c_{\bm{U}}(\Theta)}\;.
\end{eqnarray*}
 Therefore, we have that \[I_{ii}(\Theta)=\frac{1}{w_i^2}+\kappa_i^2  \frac{b(\Theta)}{c_{\bm{U}}(\Theta)}=\frac{1}{w_i^2}+\kappa_i^2\Gamma_{\bm{U}}(\Theta).\]
  Likewise, for $i\neq j$, \[I_{ij}=-\E_X\left[\frac{\partial^2 \lambda(X,\Theta)}{\partial \theta_i \partial \theta_j}\right]=\kappa_i\kappa_j c_u(\theta)\E[X]=\kappa_i\kappa_i \frac{b(\Theta)}{c_{\bm{U}}(\Theta)}=\kappa_i\kappa_i\Gamma_{\bm{U}}(\Theta)\;.\]
\end{proof}

\subsection{Appendix C: Proof of Proposition \ref{propFixPointsGen} and Corollary \ref{propFixPoints}}
\begin{proof}
The Proof of Proposition \ref{propFixPointsGen} is easy generalization from the two traits model.\\
First, assume that we are in the presence of two traits.\\
We have that  \[\mbox{det}(\Sigma)= 0\Longleftrightarrow \sigma_{11}\sigma_{22}=\sigma_{12}\sigma_{21}\Longleftrightarrow \mu_1\mu_2=1\;.\] The system in \eqref{eqn55} has non trivial solution if $\Sigma g(x,\Theta)=0$, that is, 

\begin{equation}\label{eqn:sol}
\begin{cases}
g_1(x,\theta_1,\theta_2)+\mu_2 g_2(x,\theta_1,\theta_2)&=0\\
  \mu_1g_1(x,\theta_1,\theta_2)+ g_2(x,\theta_1,\theta_2)&=0
  \end{cases}\;.
  \end{equation} We will show in the sequel that either of the  equations in \eqref{eqn:sol} can be used the characterize the solutions $(\theta_1,\theta_2)$.
Let $\ds H=c_0\mbox{exp}\left(-\ds \sum_{i=1}^n\kappa_i(\theta_{i*}-u_i)\right)$,  for  $n=2$. From the first equation in \eqref{eqn:syst}, we have
  \begin{eqnarray*}
  G(x_*,\theta_{1*},\theta_{2*})=1 &\Longleftrightarrow& b_0\mbox{exp}\left(\ds- \sum_{i=1}^n \frac{\theta_{i*}^2}{2w_i^2}-c_0 x_* \mbox{exp}\left(\ds -\sum_{i=1}^n \kappa_i(\theta_{i*}-u_i)\right)\right)=1\\
  &\Longleftrightarrow&   b_0=\mbox{exp}\left(\ds \sum_{i=1}^n \frac{\theta_{i*}^2}{2w_i^2}+c_0 x_*\mbox{exp}\left(\ds -\sum_{i=1}^n \kappa_i(\theta_{i*}-u_i)\right)\right)\\
    &\Longleftrightarrow& b_0=\mbox{exp}\left(\ds \sum_{i=1}^n \frac{\theta_{i*}^2}{2w_i^2}+ x_* H\right)\\
        &\Longleftrightarrow& \ln(b_0)= \sum_{i=1}^n \frac{\theta_{i*}^2}{2w_i^2}+ x_* H\;.
 \end{eqnarray*}
 It therefore follows that 
\begin{equation}\label{eqn:xstar}
x^*=\frac{\ds \ln(b_0)-\sum_{i=1}^n \frac{\theta_{i*}^2}{2w_i^2}}{H}\;.
\end{equation}
Next, we define  \[\ds \xi_2=\frac{1}{\nu^2}\left(\frac{1}{w_1^2}+\frac{\mu_2^2}{w_2^2}\right)+2\ln(b_0)\;.\]
Therefore, for  a solution $(x,\theta_1,\theta_2)$ of \eqref{eqn:syst}, we have 
\begin{eqnarray*}
g_1(x,\theta_1,\theta_2)+\mu_2 g_2(x,\theta_1,\theta_2)&=& -\frac{\theta_1}{w_1^2}+x\kappa_1H+\mu_2\left( -\frac{\theta_2}{w_2^2}+x\kappa_2H\right)\\
&=&-\left(\frac{\theta_1}{w_1^2}+\mu_2\frac{\theta_2}{w_2^2}\right)+xH(\kappa_1+\mu_2\kappa_2)\\
&=& -\left(\frac{\theta_1}{w_1^2}+\mu_2\frac{\theta_2}{w_2^2}\right)+\nu\left( \ln(b_0)-\frac{\theta_1^2}{2w_1^2}-\frac{\theta_2^2}{2w_2^2}\right)\\
&=& -\frac{\nu}{2w_1^2} \theta_1^2-\frac{1}{w_1^2}\theta_1-\frac{\nu}{2w_2^2} \theta_2^2-\frac{\mu_2}{w_2^2}\theta_2+\nu\ln(b_0)\;.\\
&=& -\frac{\nu}{2w_1^2} \left(\theta_1^2+\frac{2}{\nu}\theta_1\right)-\frac{\nu}{2w_2^2} \left(\theta_2^2+\frac{2\mu_2}{\nu}\theta_2\right)+\nu\ln(b_0)\;.\\
&=&-\frac{\nu}{2w_1^2} \left(\theta_1+\frac{1}{\nu}\right)^2+\frac{\nu}{2w_1^2} \frac{1}{\nu^2}-\frac{\nu}{2w_2^2} \left(\theta_2+\frac{\mu_2}{\nu}\right)^2\\
&+&\frac{\nu}{2w_2^2}\frac{\mu_2^2}{\nu^2}+\nu\ln(b_0)\;.\\
&=&-\frac{\nu}{2w_1^2} \left(\theta_1+\frac{1}{\nu}\right)^2-\frac{\nu}{2w_2^2} \left(\theta_2+\frac{\mu_2}{\nu}\right)^2+\frac{1}{2w_1^2\nu}\\
 &+&\frac{\mu_2^2}{2w_2^2\nu}+\nu\ln(b_0)\;.\\
\end{eqnarray*}
Dividing the latter by $\frac{\nu}{2}$, we have that 
\begin{eqnarray*}
g_1(x,\theta_1,\theta_2)+\mu_2 g_2(x,\theta_1,\theta_2)=0&\Longleftrightarrow&-\frac{1}{w_1^2} \left(\theta_1+\frac{1}{\nu}\right)^2-\frac{1}{w_2^2} \left(\theta_2+\frac{\mu_2}{\nu}\right)^2+\frac{1}{w_1^2\nu^2}\\
 &+&\frac{\mu_2^2}{w_2^2\nu^2}+2\ln(b_0)=0\;.\\
 &\Longleftrightarrow&\frac{1}{w_1^2} \left(\theta_1+\frac{1}{\nu}\right)^2\frac{1}{w_2^2}\left(\theta_2+\frac{\mu_2}{\nu}\right)^2-\xi_2=0\\
\end{eqnarray*}
Clearly,  if $\xi_2<0$, there is no solution to $g_1(x,\theta_1,\theta_2)+\mu_2 g_2(x,\theta_1,\theta_2)=0$. \\
If $\xi_2=0$, then \[g_1(x,\theta_1,\theta_2)+\mu_2 g_2(x,\theta_1,\theta_2)=0\Longleftrightarrow (\theta_1,\theta_2)=\left( -\frac{1}{\nu},-\frac{\mu_2}{\nu}\right)\;.\]
If $\xi_2>0$, we let 
\[a=w_1\sqrt{\xi_2}, \quad  b=w_2\sqrt{\xi_2}\;.\]
It follows that \[g_1(x,\theta_1,\theta_2)+\mu_2 g_2(x,\theta_1,\theta_2)=0\Longleftrightarrow \frac{\left(\theta_1+\frac{1}{\nu}\right)^2}{a^2}+ \frac{\left(\theta_2+\frac{\mu_2}{\nu}\right)^2}{b^2}=1\;.\]
  That is,  the ellipse centered at $(\theta_{1}^0,\theta_{2}^0)=\left(-\frac{1}{\nu},-\frac{\mu_2}{\nu}\right)$ with respective major and minor axis lengths 
 $a$ and $b$. Similarly to above, define  \[\ds \xi_*=\frac{1}{\nu_*^2}\left(\frac{\mu_1}{w_1^2}+\frac{1}{w_2^2}\right)+2\ln(b_0), \quad \mbox{where  $\nu_*=\mu_1\kappa_1+\kappa_2$}\;.\]
If $\xi_*>0$, then  \[\mu_1g_1(x,\theta_1,\theta_2)+ g_2(x,\theta_1,\theta_2)=0\Longleftrightarrow \frac{\left(\theta_1+\frac{\mu_1}{\nu_*}\right)^2}{a_*^2}+ \frac{\left(\theta_2+\frac{1}{\nu_*}\right)^2}{b_*^2}=1\;.\] That is,   the ellipse centered at $(\theta_{1*}^0,\theta_{2*}^0)=\left(-\frac{\mu_1}{\nu_*},-\frac{1}{\nu_*}\right)$ with respective major and minor axis lengths 
\[a_*=w_1\sqrt{\xi_*}, \quad  b_*=w_2\sqrt{\xi_*}\;.\]
We observe that the two ellipses  are the same: \\
\noindent 1) They have the identical centers.  Indeed, we have $\nu=\mu_2\nu_*$ and $\ds \frac{1}{\nu}=\frac{1}{\mu_2\nu_*}=\frac{\mu_1}{\nu_*}$ since $\mbox{det}(\Sigma)= 0$ implies that  $\mu_1\mu_2=1$.  Likewise, we have $\ds \frac{1}{\nu_*}=\frac{\mu_2}{\nu}$.\\
\noindent This proves that $(\theta_{1}^0,\theta_{2}^0)=(\theta_{*1}^0,\theta_{*2}^0)$, that is, the two centers are identical.\\
2) They have the same parameters.  Indeed, we have 
\begin{eqnarray*}
\frac{1}{\nu^2}\left(\frac{1}{w_1^2}+\frac{\mu_2^2}{w_2^2}\right)&=&\frac{1}{\mu_2^2\nu_*^2}\left(\frac{1}{w_1^2}+\frac{\mu_2^2}{w_2^2}\right)\\
&=&\frac{1}{\nu_*^2}\left(\frac{1}{\mu_2^2w_1^2}+\frac{1}{w_2^2}\right)\\
&=&\frac{1}{\nu_*^2}\left(\frac{1}{w_2^2}+\frac{\mu_1^2}{w_1^2}\right), \quad \mbox{since $\mu_1\mu_2=1$.}
\end{eqnarray*}
This implies that $\xi_2=\xi_*$ and thus  $a=a_*$ and $b=b_*$.  \\
To generalize, we note that $\Sigma g(x,\Theta)=0$ implies that  for given $1\leq j\leq n$, we 
\[\sum_{i=1}^n\sigma_{ji}g_i(x,\Theta)=0.\]
Since we assume that $\sigma_{ii}\neq 0$, without loss of generality, let $j=1$. Then we have 
\[\sum_{i=1}^n\sigma_{ji}g_i(x,\Theta)=0 \Longleftrightarrow g_1(x,\Theta)+\mu_{12}g_2(x,\Theta)+\cdots +\mu_{1n}g_n(x,\Theta)\;. \]
Rearranging the terms and completing the squares, we obtain the result as announced. 

\end{proof}


 \end{document}